\documentclass[fleqn,usenatbib]{mnras}

\usepackage{newtxtext,newtxmath}

\usepackage[T1]{fontenc}

\DeclareRobustCommand{\VAN}[3]{#2}
\let\VANthebibliography\thebibliography
\def\thebibliography{\DeclareRobustCommand{\VAN}[3]{##3}\VANthebibliography}

\usepackage{graphicx}	
\usepackage{amsmath}	
\usepackage{color}
\usepackage{enumitem}
\usepackage{hyperref}
\usepackage{verbatim}
\usepackage{mathrsfs}
\usepackage[all]{hypcap} 
\usepackage{afterpage}    
\usepackage{marginnote}
\usepackage{multirow}
\usepackage{lineno}
\usepackage{float}
\usepackage{siunitx}

\title[Magnetic field in the Presence of Stellar Feedback]{Magnetic Fields and Velocity Gradients in L1551: The Role of Stellar Feedback}

\author[Liu, Hu \& Lazarian]{
Mingrui Liu$^{1}$,
Yue Hu$^{2,3}$\thanks{E-mail: yue.hu@wisc.edu},
Alex Lazarian$^{3}$\thanks{E-mail: alazarian@facstaff.wisc.edu}
\\
$^{1}$Department of Applied Physics and Applied Mathematics, Columbia University, New York, NY, 10076, USA\\
$^{2}$Department of Physics, University of Wisconsin-Madison, Madison, WI, 53706, USA\\
$^{3}$Department of Astronomy, University of Wisconsin-Madison, Madison, WI, 53706, USA\\
}

\date{Accepted XXX. Received YYY; in original form ZZZ}

\pubyear{2023}

\begin{document}
\label{firstpage}
\pagerange{\pageref{firstpage}--\pageref{lastpage}}
\maketitle

\begin{abstract}
Magnetic fields play a crucial role in star formation, yet tracing them becomes particularly challenging, especially in the presence of outflow feedback in protostellar systems. We targeted the star-forming region L1551, notable for its apparent outflows, to investigate the magnetic fields. These fields were probed using polarimetry observations from the Planck satellite at 353 GHZ/849 $\mu$m, the SOFIA/HAWC+ measurement at 214 $\mu$m, and the JCMT/SCUPOL 850 $\mu$m survey. Consistently, all three measurements show that the magnetic fields twist towards the protostar IRS 5.  Additionally, we utilized the Velocity Gradients Technique (VGT) on the $^{12}$CO (J = 1-0) emission data to distinguish the magnetic fields directly associated with the protostellar outflows. These were then compared with the polarization results. Notably, in the outskirts of the region, these measurements generally align. However, as one approaches the center of IRS 5, the measurements tend to yield mostly perpendicular relative orientations. This suggests that the outflows might be dynamically significant from a scale of approximately $\sim0.2$~pc, causing the velocity gradient to change direction by 90 degrees. Furthermore, we discovered that the polarization fraction $p$ and the total intensity $I$ correlate as $p \propto I^{-\alpha}$. Specifically, $\alpha$ is approximately $1.044\pm0.06$ for SCUPOL and around $0.858\pm0.15$ for HAWC+. This indicates that the outflows could significantly impact the alignment of dust grains and magnetic fields in the L1551 region.
\end{abstract}


\section{Introduction}
The interstellar medium (ISM; \citealt{1981MNRAS.194..809L,1995ApJ...443..209A,2004ARAA..42..211E}) is permeated by magnetic fields, which display a wide range of complexity on different scales (\citealt{Crutcher12, BG15,2016A&A...586A.136P,2022ApJ...941...92H}). These fields have a major impact on the behavior of protostellar material, from the filaments on the parsec scale in molecular clouds to the sub-AU regions around protostars (\citealt{1978prpl.conf..243F,2007ApJ...670.1198M,2010Sci...330.1209C,2013ApJ...769L..15S}; \citealt{2021ApJ...916...78C}). They can impede or promote the gravitational collapse of molecular clouds (\citealt{1981MNRAS.194..809L,1987ARA&A..25...23S,2004RvMP...76..125M,2007ARA&A..45..565M,Crutcher12,2012ApJ...761..156F,2022A&A...668A.147H}) and potentially affect powerful molecular outflows from protostellar systems \citep{2017ApJ...847..104O,2019FrASS...6....7K,2023ApJ...952...29K}.

The exact mechanism of protostellar outflows is still unclear; however, magnetic fields are thought to be a key component, as they can accelerate the transfer of angular momentum during protostar formation \citep{2016ARA&A..54..491B}. Some studies suggest that outflow feedback might modify the magnetic field structure on scales of around 1000 AU \citep{2014ApJS..213...13H,2019FrASS...6....3H,2022ApJ...941...81X}; however, others have indicated that the outflow geometry and the plane-of-the-sky (POS) magnetic field may not be interconnected, displaying random relative orientation on larger scales \citep{2022MNRAS.515.1026P}. This discrepancy has yet to be resolved. The magnetic field is primarily inferred from the polarized thermal emission of interstellar dust grains, which tend to align their semi-major axes perpendicular to local magnetic fields due to radiative torques (RATs; \citealt{2007MNRAS.378..910L,BG15}). Dust grains may retain this alignment in both diffuse parts in the ISM and also the dense regions with a volume density of up to $10^4~{\rm cm^{-3}}$ (\citealt{2008MNRAS.387..797H,2019MNRAS.482.2697S}). Meanwhile, however, the diffuse molecular outflows are typically traced by diffuse $^{12}$CO and $^{13}$CO corresponding to the critical density of
$10^2$ and $10^3~{\rm cm^{-3}}$, respectively, when only collisional (with H$_2$) de-excitation and spontaneous radiative de-excitation at kinetic temperature $\sim10~$K are considered \citep{1978ApJ...222..881G,1998AJ....116..336N,2011piim.book.....D}. Such differences imply that the dust polarization measurements may reflect the superposition of magnetic fields in both the diffuse and dense regions, not necessarily those explicitly associated with the molecular outflows.

To directly trace the magnetic field associated with outflows, the Velocity Gradient Technique (VGT; \citealt{2017ApJ...835...41G,2017ApJ...837L..24Y,2023MNRAS.524.2994H}) offers a promising way. VGT is founded on the understanding that in MHD turbulence, eddies elongate along the magnetic field, exhibiting anisotropy \citep{GS95,LV99}. The velocity gradient of these turbulent eddies acts as a detector for this anisotropy and can, therefore, be used to trace the magnetic fields. The anisotropy, in turn, is imprinted on spectroscopic observations (\citealt{2000ApJ...537..720L,2017MNRAS.464.3617K,2021ApJ...915...67H,2023MNRAS.519.3736H}). Thus, VGT can adeptly utilize molecular emission lines to trace magnetic fields. However, the presence of outflow feedback may alter turbulence properties (\citealt{2022MNRAS.511..829H}) and impose additional velocity gradients that are not associated with MHD turbulence. This effect is expected to be scale-dependent as outflows are typically weak at large scales, but this relationship has not yet been explored.

To study the magnetic field and velocity gradient properties amidst outflow, it is vital to obtain multi-scale and multi-frequency magnetic field measurements. These are now readily available through the use of advanced polarimetry on dust emission, such as the \textit{Planck} satellite 353 GHz/849 $\mu$m survey (\citealt{2020A&A...641A..12P}), the JCMT/SCUPOL survey at 850 $\mu$m (\citealt{2009ApJS..182..143M}), and the SOFIA/HAWC+ survey at 214 $\mu$m (\citealt{2018JAI.....740008H}). These instruments have been utilized for observing the molecular cloud L1551 (\citealt{1980ApJ...239L..17S,1981ApJS...45..121S,2010ApJ...718.1019Y,2016ApJ...826..193L}), a low-mass star-forming cloud (M $\approx 50$ M$_\odot$) exhibiting substantial bipolar molecular outflow from young stellar objects L1551 IRS 5 (hereafter IRS 5) and L1551 NE (hereafter NE).  Given that L1551 is located approximately 160 pc away, SCUPOL and SOFIA/HAWC+ offer high spatial resolutions of $10''$ and $18.2''$ respectively, capably resolving scales of 0.008 pc and 0.014 pc associated with stronger outflows. Importantly, the measurements from these two distinct wavelengths are expected to trace the magnetic field associated with different gas/dust temperatures. Additionally, we apply VGT to $^{\rm 12}$CO (J = 1-0) emission lines, while its application to $^{\rm 13}$CO (J = 1-0) has been investigated \citep{2019NatAs...3..776H}. As $^{\rm 12}$CO is a common tracer for protostellar outflows \citep{1980ApJ...239L..17S,2006ApJ...649..280S,2007prpl.conf..215B, 2016ARA&A..54..491B}, the magnetic fields inferred by VGT from the molecular gas emission are primarily associated with outflows in the star-forming region. Comparison of results from VGT and polarimetry will yield unique insights into the properties of velocity gradients in the presence of outflow feedback.

The paper is structured as follows: \S~\ref{sec:ob} provides an overview of the observational data used in our study. In \S~\ref{sec:method}, we dive into the VGT, discussing its theoretical foundation in terms of MHD turbulence and the software framework. Our principal findings concerning the magnetic fields in L1551, as derived from polarization measurements and VGT, are revealed in \S~\ref{sec:res}. Subsequently, \S~\ref{sec:dis} offers a discussion of the various magnetic field measurements. Our conclusions are drawn in \S~\ref{sec:con}.


\section{Observations}
\label{sec:ob}
\subsection{$^{12}$CO emission}
The $^{12}$CO (J = 1-0, 115.27120 GHz) emission line of L1551 was observed by the Nobeyama 45m radio telescope from December 2007 to May 2008 (\citealt{2010ApJ...718.1019Y}). At 115 GHz, the telescope has a half-power beam width (HPBW) of 15$''$. The region was mapped into a 45$'\times$45$'$ sample with a 7.$''$5 pixel grid size. The velocity resolution of the data is 37.8 kHz in frequency or 0.098 $\rm km$ $\rm s^{-1}$ in velocity, along with an RMS noise level of 1.23 K in $T_{\rm mb}$ (\citealt{2016ApJ...826..193L}). We applied the VGT to the cube with a grid resolution of 7.$''$5. All velocity channels within the velocity range from -9.9 $\rm km$ $\rm s^{-1}$ to 14.9 $\rm km$ $\rm s^{-1}$, in which the cloud’s emission is concentrated (see Appendix ~\ref{app:sm} for the gas emission spectrum), are considered in VGT calculation.

\subsection{Polarized dust emission}
We adopt the following polarized dust emission data: Planck 353 GHz data from the 3rd Public Data Release (\citealt{2020A&A...641A..12P}), SOFIA/HAWC+ observation, and JCMT/SCUPOL survey (\citealt{2009ApJS..182..143M}). 

\textbf{Planck:} the Planck polarization has beam resolution $\approx$ 5$'$. The magnetic field orientation $\phi_{\rm B}=\phi+\pi/2$ is derived from the polarization angle $\phi$, which is calculated by the Stokes $Q$ and $U$:
        \begin{equation}
        \begin{aligned}
           \phi&=\frac{1}{2}\arctan(-U,Q),\\
        \end{aligned}
        \end{equation}
where the negative sign in $-U$ converts the angle originally in the HEALPix system to the IAU system. Note the $\pi$ periodicity in radian denoted by the arc-tangent function. The polarization fraction $p$ is defined as:
\begin{equation}
    \begin{aligned}
        p&=\frac{\sqrt{Q^2+U^2}}{I}
    \end{aligned}
\end{equation}
where I is the intensity of dust emission.
However, the polarization needs to be further debaised to correct for the instrumental error. We employ the common debiasing recipe as in \cite{1974ApJ...194..249W} and \cite{2019ApJ...880...27P}:
\begin{equation}
        \begin{aligned}
           p_d&=\frac{\sqrt{Q^2+U^2-\frac{1}{2}(\delta Q^2+\delta U^2)}}{I},
        \end{aligned}
        \end{equation}
$\delta Q$ and $\delta U$ are the uncertainty in the Stokes $Q$ and $U$. 

\textbf{SOFIA/HAWC+:} The observation was carried out in band E (214 $\mu$m) with an HPBW of $\sim$ 18.$^{''}$2 in November 2021. Similar to Planck, the magnetic field orientation is derived from Stokes $Q$ and $U$. We filter the data to keep only pixels that satisfy: i) $p/\delta p>3$; ii) $I/\delta I>4$, where $\delta p$ and $\delta I$ are the uncertainty of polarization fraction and intensity, respectively. In addition, it is observationally rare to obtain a polarization fraction higher than 30\% in diffuse ISM \citep{2022MNRAS.512.1985F}. For example, \cite{2020A&A...641A...3P} reported that the polarization fraction across the full sky has a maximum of approximately 22$\%$. Therefore, pixels with a polarization fraction larger than 25$\%$ are blanked out.


\textbf{JCMT/SCUPOL}: The data used in this work was retrieved from the SCUBA Polarimeter Legacy Catalogue\footnote{www.cadc-ccda.hia-iha.nrc-cnrc.gc.ca/en/community/scupollegacy} with an angular resolution of 10$''\times$10$''$ 
at 850 $\mu$m \citep{2009ApJS..182..143M}. The data was processed by the SCUPOL team to achieve the following criteria: i) $p/\delta p$ > 2; ii) $\delta p$ < 4\% ; iii) I > 0.

 

\section{Methodology}
\label{sec:method}
\subsection{Essential of MHD turbulence and the Outflow Effect}
\textbf{Anisotropy of MHD turbulence:} In decades, our understanding of MHD turbulence has rapidly changed. Many studies revealed that turbulence in the presence of magnetic fields is isotropic rather than anisotropic. Especially the cornerstone theoretical frame developed in \cite{GS95} for trans-Alfv\'enic turbulence (i.e., the Alfv\'en Mach number M$_A\sim1$) found the turbulent eddies are elongating along the magnetic fields, exhibiting the scaling relation:
\begin{equation}
k_\parallel\propto k_\bot^{2/3},
\end{equation}
 where $k_\parallel$ and $k_\bot$ represent the components of the wavevector parallel and perpendicular to the mean magnetic field, respectively. Also, the "critical balance" condition, i.e., the cascading time ($k_\bot v_l$)$^{-1}$ equals the wave periods ($k_\parallel v_A$)$^{-1}$, and Kolmogorov-type turbulence, i.e., $v_l\propto l^{1/3}$, were considered. Nonetheless, it should be noted that the "critical balance" condition considered in \cite{GS95} was based on a global reference frame, where the direction of wavevectors is defined relative to the mean magnetic field. 

 For the purpose of tracing magnetic fields, it is important to find the anisotropy in the local reference frame defined relative to the magnetic field passing through an eddy at scale $l$. This was later achieved by \cite{LV99} in the study of turbulent reconnection. According to \cite{LV99}, turbulent reconnection of the magnetic field, takes place over just one eddy turnover time, naturally giving the "critical balance" in the local frame: $v_{l,\bot}l_\bot^{-1}\approx v_Al_\parallel^{-1}$, where $l_\bot$ and $l_\parallel$ represent the perpendicular and parallel scales of eddies with respect to the local magnetic field, respectively. Also, the motion of eddies perpendicular to the local magnetic field direction adheres to the Kolmogorov law (i.e., $v_{l,\bot}\propto l_\bot^{1/3}$), since this is the direction in which the magnetic field offers minimal resistance. The scale-dependent anisotropy scaling is then given by:
\begin{equation}
\label{eq.lv99}
 l_\parallel= L_{\rm inj}(\frac{l_\bot}{L_{\rm inj}})^{\frac{2}{3}}{\rm M}_A^{-4/3},~M_{\rm A}\le 1,
 \end{equation}
where $l_\bot$ and $l_\parallel$ represent the perpendicular and parallel scales of eddies with respect to the local magnetic field, respectively. $L_{\rm inj}$ denotes the turbulence injection scale and ${\rm M}_A=v_{\rm inj}/v_A$. On the other hand, the scaling of turbulent velocity and its corresponding gradient are:
\begin{equation}
\label{eq.5}
        \begin{aligned}
           &v_{l,\bot}  =  v_{\rm inj}(\frac{l_{\bot}}{L_{\rm inj}})^{1/3}M_{\rm A}^{1/3},~M_{\rm A} \le 1,\\
           &\nabla v_l\propto \frac{v_{l,\bot}}{l_{\bot}}=\frac{v_{\rm inj}}{L_{\rm inj}}M_{\rm A}^{1/3}(\frac{l_{\bot}}{L_{\rm inj}})^{-2/3},~ M_{\rm A} \le 1.
        \end{aligned}
\end{equation}
Here the velocity gradient is dominated by the perpendicular component, due to the anisotropy (i.e., $l_\parallel\gg l_\bot$). One can therefore derive the local magnetic field orientation from the velocity gradient. 

\textbf{Outflow Effects}: Our earlier discussion presupposes the dominance of MHD turbulence. Nonetheless, within star-forming regions, outflow feedback might alter the properties of velocity gradients.
Outflow feedback has been shown to substantially change fluid velocity statistics (\citealt{2022MNRAS.511..829H}) and is anticipated to mirror the effects of inflows on velocity gradients, i.e., causing the orientation of the velocity gradient to shift from being perpendicular to the magnetic fields to being aligned with them \citep{HLY20}. Since $^{12}$CO only resolves outflows in the L1551 region \citep{2006ApJ...649..280S}, we focus our analysis on the shift of velocity gradients by outflows rather than inflows. It is well-documented that outflows are expelled from star-forming sites at high velocities (\citealt{1980ApJ...239L..17S}). At a protostar's core, gas velocity peaks and diminishes towards its periphery. As a result, additional velocity gradients not attributed to turbulence might emerge, pointing from the center to the outskirts. These outflow-induced gradients are highly scale-dependent because of their correlation with the significance of outflows. On smaller scales proximate to protostars, outflows significantly overshadow turbulence. Magnetic fields in this case are expected to follow the outflows. Thus, the velocity gradient is primarily dominated by the potent outflows, being parallel to the magnetic fields. Conversely, on broader scales distant from protostars, weakened outflows allow turbulence-induced velocity gradients to take precedence, rendering the gradients perpendicular to the magnetic fields. A transition may exist wherein velocity gradients gradually shift from parallel to perpendicular alignment with the magnetic fields as one moves further from the protostar's center, contingent upon the comparative prominence of outflows versus turbulence.

\subsection{VGT pipeline}
\textbf{Velocity caustics effect:} Employing VGT requires velocity information to be extracted at each pixel of observed images. For such purpose, we treat the Doppler-shifted emission line data into maps with thin velocity channels. This is inspired by the velocity caustics effect (\citealt{2000ApJ...537..720L,2017MNRAS.464.3617K}), which indicates that in observations, physical (density or intensity) structures are sampled into different channels with respect to their LOS velocities. In turbulent ISM, the intensity structures in each channel are distorted. Such distortion becomes more significant as the channel width gets narrower. Velocity fluctuations will eventually dominate over intensity fluctuations on $\Delta v < \sqrt{\delta(v^2)}$: the channel width $\Delta v$ is smaller than the velocity dispersion $\sqrt{\delta(v^2)}$. In such thin channels, calculating the gradient of intensity amplitudes would in fact generate gradient information of the velocity fluctuations. More details about the thin channel as well as the Position-Position-Velocity (PPV) theory can be found in \citet{2000ApJ...537..720L,2017MNRAS.464.3617K,2023MNRAS.524.2994H}.

\begin{figure}
\label{fig.outflows}
    \centering
    \includegraphics[width=1\linewidth]{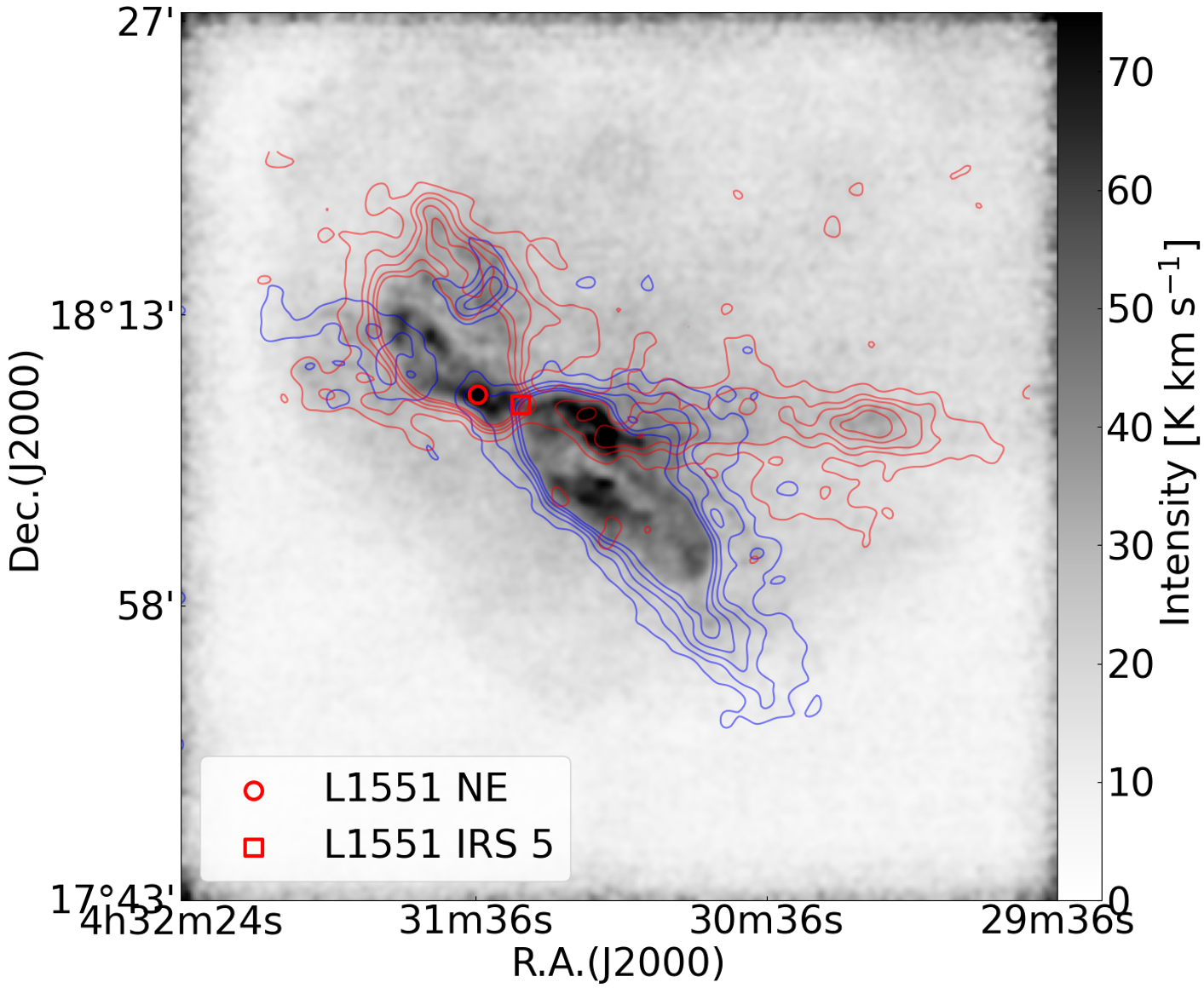}
    \caption{Integrated intensity map of the $^{\rm 12}$CO (1-0) emission in L1551 with the redshifted and blueshifted components marked by contours in respective colors. Each intensity contour is separated from adjacent levels by 25 k $\rm km$ $\rm s^{-1}$. The redshifted emission is marked with contour levels [25, 50, 70, 100, 125] k $\rm km$ $\rm s^{-1}$ with a velocity range of $v_{\rm LSR}$ = 7.9 $\sim$ 17.4 $\rm km$ $\rm s^{-1}$, and the blueshifted emission has contour levels [20, 45, 70, 95, 120] k $\rm km$ $\rm s^{-1}$, residing in $v_{\rm LSR}$ = -3.5 $\sim$ 5.5 $\rm km$ $\rm s^{-1}$. Two protostar systems in the region, L1551 NE (red circle) and L1551 IRS 5 (red square), have also been highlighted.}
\end{figure}

\begin{figure*}
\label{fig.bfields}
    \centering
    \includegraphics[width=1.003\linewidth]{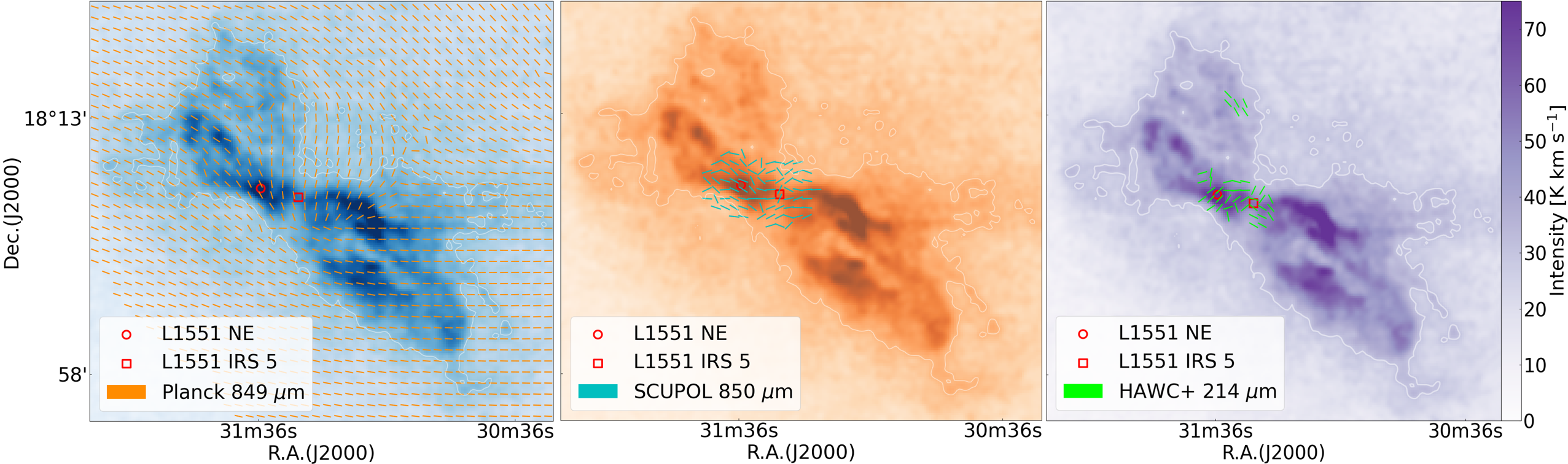}
    \caption{Magnetic field maps inferred from Planck (left), SCUPOL (middle), and HAWC+ (right). The magnetic field orientations are represented by colored line segments, overlaid upon the integrated intensity map of the $^{12}$CO emission. SCUPOL and HAWC+ only surveyed a small portion of the L1551 region, thus the maps are zoomed in to the main intensity structure for better illustration, and so does every figure that contains results from the two surveys in the following sections.}
\end{figure*}

\begin{figure*}
\label{fig.p_planck}
    \centering
    \includegraphics[width=1\linewidth]{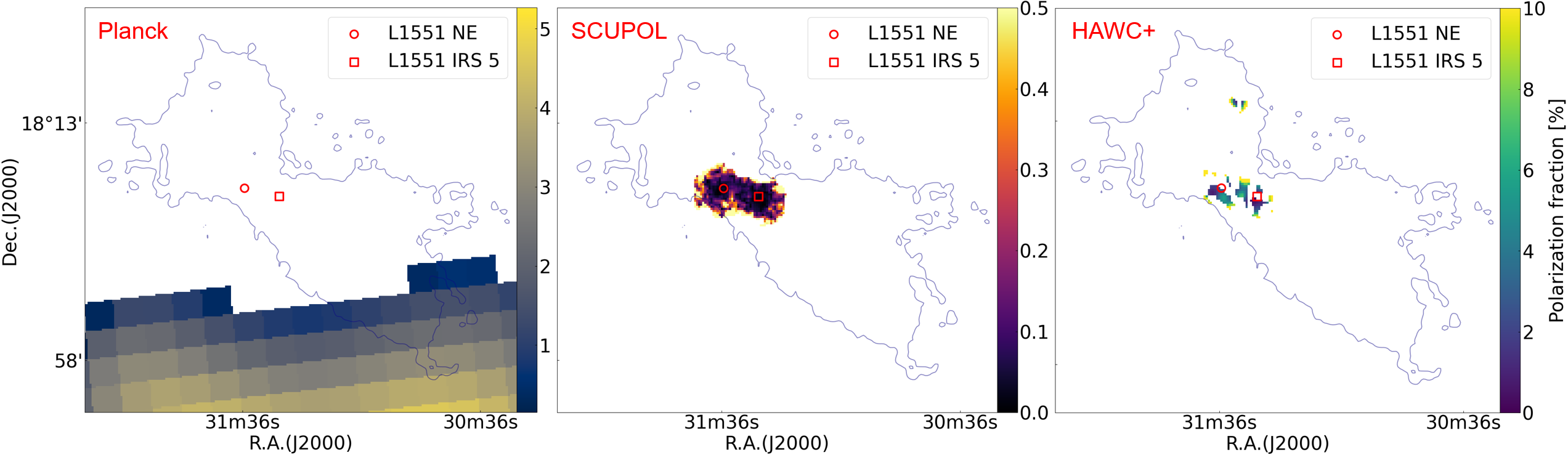}
    \caption{Polarization fraction per pixel in the Planck (left), SCUPOL (middle), and HAWC+ (right) data. Only pixels that satisfy both the following criteria are shown: i) p > 0\%; ii) p < 25\%.}
\end{figure*}

\textbf{VGT:} The software workflow for VGT used in this work follows the procedure introduced in \cite{2022MNRAS.511..829H} which is summarized below: each individual thin channel map $\rm Ch_{i}(x, y)$  $(i = 1, 2, \cdots, n_v)$ is convolved with the 3$\times$3 Sobel convolutional kernels\footnote{
\begin{equation}
       \begin{aligned}
                  &G_x = 
        \begin{pmatrix}
         -1 & 0 & +1\\ 
         -2 & 0 & +2\\
         -1 & 0 & +1
        \end{pmatrix},
        G_y =
        \begin{pmatrix}
         -1 & -2 & +1\\ 
        0 & 0 & 0\\
         +1 & +2 & +1
        \end{pmatrix}\\
        \end{aligned}
\end{equation}}:
\begin{equation}
        \begin{aligned}
           \nabla_{x} {\rm Ch_i} (x,y) =  \textit{$G_{\rm x}$} * \textit{${\rm Ch_i} (x,y)$}, \\
           \nabla_{y} {\rm Ch_i} (x,y)  =  \textit{$G_{\rm y}$} * \textit{${\rm Ch_i} (x,y)$},
        \end{aligned}
        \end{equation}
the asterisks denote convolutions. $\nabla_{x} {\rm Ch_i} (x,y)$ and $\nabla_{y} {\rm Ch_i} (x,y)$ are the gradient components in each thin channel maps along the x and y axis. They are used to calculate the overall pixelized map $\psi_{\rm g}^{\rm i} (x,y)$ of the gradients:
\begin{equation}
       \begin{aligned}
&\psi{\rm ^i_g} (x,y)=\tan^{-1}(\frac{\nabla{}{}_{y} {\rm Ch_i} (x,y)}{\nabla_{x} {\rm Ch_i} (x,y)}).
\end{aligned}
\end{equation}
From here, only pixels with an intensity less than three times the RMS noise are kept towards the following steps. The resulting $\psi_{\rm g}^{\rm i}$ is then processed with the sub-block averaging method (\citealt{2017ApJ...837L..24Y}): i) the pixelized map is divided into rectangular sub-blocks with an optimal size of 20$\times$20 pixels (verified both empirically and numerically: \citealt{2018ApJ...853...96L,2021ApJ...911...37H}); ii) a histogram is produced for the mean gradient angle within each sub-block; iii) the histogram is fitted with a Gaussian distribution, the peak value of which is then defined as the most probable orientation of the gradient in that sub-block. Now the averaged gradient angle map $\psi_{\rm gs}^{\rm i} (x,y)$ constructs the pseudo-Stokes parameters $Q_g$ and $U_g$:
\begin{equation}
\begin{aligned}
    &Q_{\rm g} (x,y)  =  \sum^{n_{\rm v}}_{{\rm i=1}} {\rm Ch_i} (x,y) \cos(2\psi{\rm ^i_{gs}} (x,y)),\\
    &U_{\rm g} (x,y)  =  \sum^{n_{\rm v}}_{{\rm i=1}} {\rm Ch_i} (x,y) \sin(2\psi{\rm ^i_{gs}} (x,y)),\\
    &\psi_{\rm g} = \frac{1}{2}\tan^{-1}(\frac{U_{\rm g}}{Q_{\rm g}}),
\end{aligned}
\end{equation}
thus the POS magnetic field angle probed by the VGT $\psi_{\rm B}$ can be inferred from the pseudo-polarization angle $\psi_{\rm g}$ as $\psi_{\rm B} = \psi_{\rm g} + \pi/2$.



\begin{figure}
\label{fig.p_dis_planck}
    \centering
    \includegraphics[width=1\linewidth]{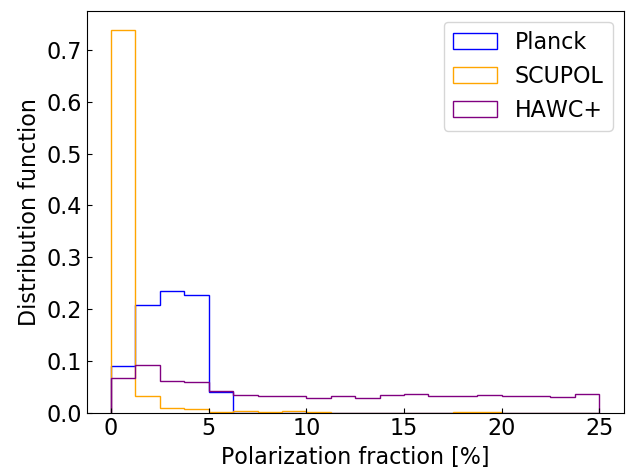}
    \caption{Histogram of polarization fraction for Planck (blue line), SCUPOL (orange line), and HAWC+ (purple line).}
\end{figure}

\begin{figure*}
\label{fig.p_hc+}
    \centering
    \includegraphics[width=1\linewidth]{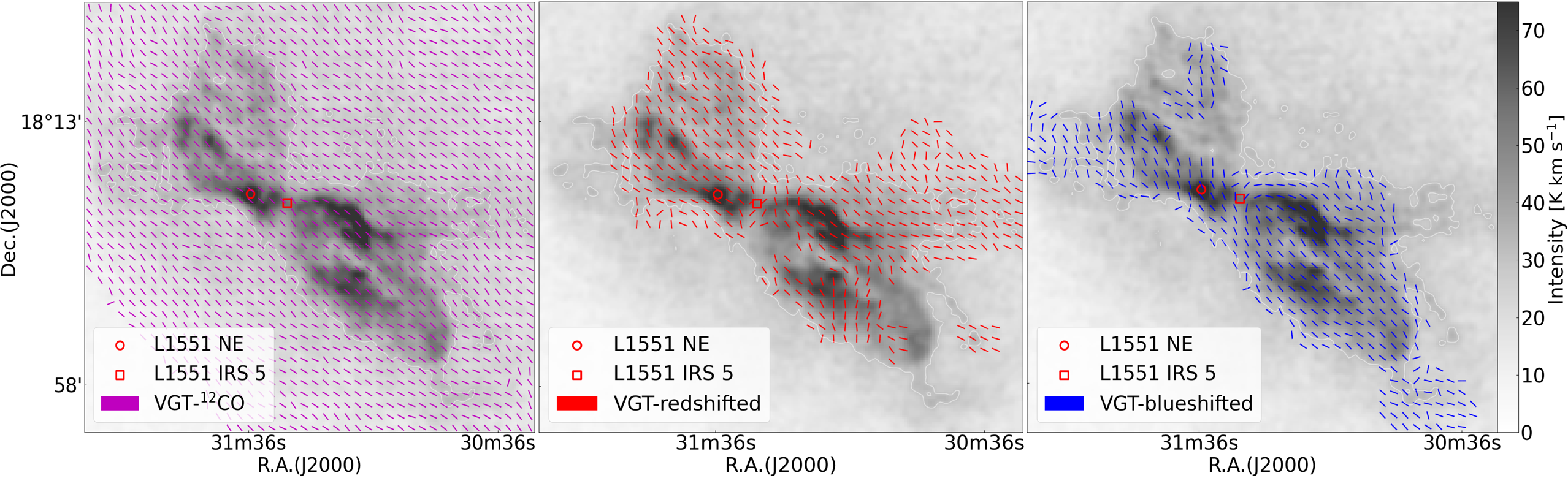}
    \caption{Magnetic field orientations of total $^{12}$CO emission (left), redshifted emission (middle), and blueshifted emission (right) obtained with VGT.}
\end{figure*}

\begin{figure*}
\label{fig.I-P_s}
    \centering
    \includegraphics[width=1\linewidth]{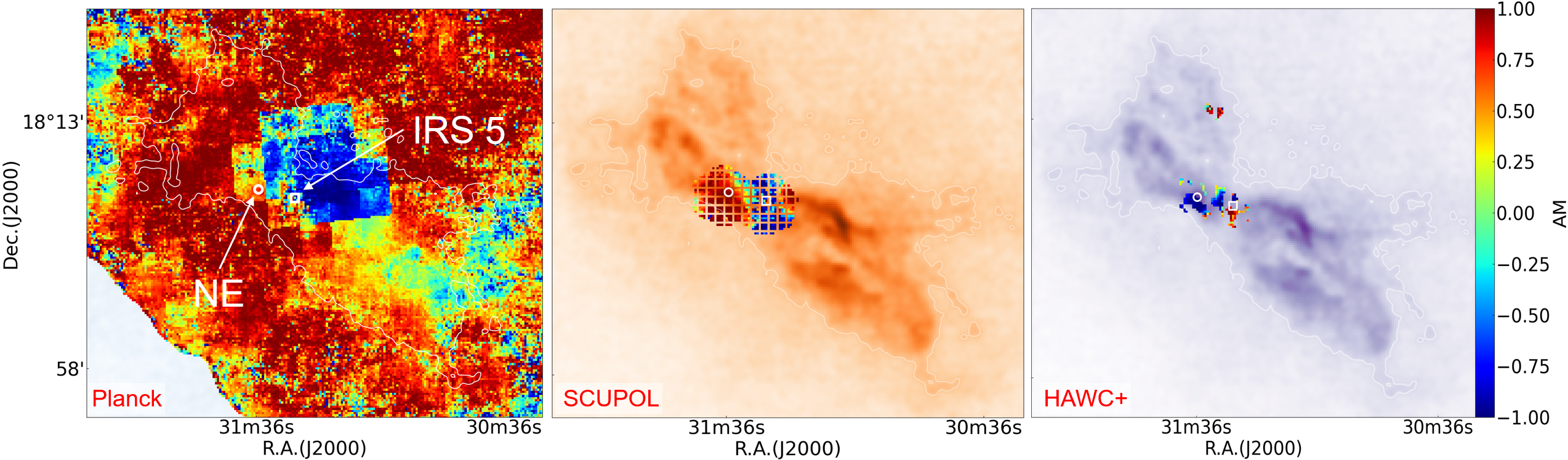}
    \caption{Spatial distributions of AM between VGT measurements on total $^{12}$CO emission and Planck (left), SCUPOL (middle), and HAWC+ (right). The positions of L1551 NE (white circle) and L1551 IRS 5 (white square) have been highlighted on each map. To resolve the inconsistency in the spatial resolutions of the polarimetry surveys, a Gaussian filter was applied to each of the polarimetry data so that every resulting map has the same 20$\times$20 pixel resolution. The graphs are different in shape with the vectors in Fig.~\ref{fig.p_planck}, which are averaged over several pixels for visualization purposes, while each map here was produced by calculating AM at every pixel.}
\end{figure*}
\section{Results}
\label{sec:res}
\subsection{Molecular outflows}
Fig.~\ref{fig.outflows} reproduced the bipolar molecular outflows observed in the L1551 region. As detailed by \cite{2010ApJ...718.1019Y}, the LSR velocities of the red- and blue-shifted components span from 7.9 $\rm km$ $\rm s^{-1}$ to 17.4 $\rm km$ $\rm s^{-1}$ and -3.7 $\rm km$ $\rm s^{-1}$ to 5.5 $\rm km$ $\rm s^{-1}$, respectively. These outflows are almost symmetric around L1551 IRS 5, the binary protostar system located at the intersection of the two components. The blue-shifted flow stretches southwest, while its red-shifted counterpart spans northeastward, with an outlying structure that extends westward for approximately 20$'$. These redshifted outflows might originate from a newly formed star near HH 102, a Herbig–Haro diffuse reflection nebula situated 5$'$ west of L1551 IRS 5 (\citealt{1976AJ.....81..320S,2006ApJ...649..280S}). Nevertheless, other studies, like \cite{2006ApJ...645..357M}, suggest that the east-west outflow's driving source is one of the stars in the multi-star system L1551 NE (\citealt{2002AJ....124.1045R,2010ApJ...718.1019Y}).

\subsection{Magnetic fields inferred from polarimetry}
Fig.~\ref{fig.bfields} presents the POS magnetic field directions inferred from three polarization measurements, including Planck (849 $\mu $m, HPBW$\sim5'$), JCMT/SCUPOL (850 $\mu$m, HPBW$\sim10''$), and HAWC+ (214 $\mu$m, HPBW$\sim18.2''$). The Planck map reveals that the magnetic fields appear to follow the northeastern $^{12}$CO intensity structure, which corresponds to the positions of the red-shifted outflows (see Fig.~\ref{fig.outflows}). On the contrary, in the southwest area, the magnetic fields are seen to cross the $^{12}$CO intensity structure, which is associated with the positions of the blue-shifted outflows (see Fig.~\ref{fig.outflows}). In particular, at the center of L1551, i.e. around the IRS 5 protostar
, the magnetic fields are significantly twisted.

We observed that Planck's polarization fraction around the cloud is negative after debiasing, as seen in Fig.~\ref{fig.p_planck}, indicating a considerable depolarization effect. This could be due to (1) turbulent magnetic fields on the POS and (2) significant variation of magnetic fields along the LOS. The low polarization fraction implies a very low signal-to-noise ratio. However, this could be reduced by increasing the resolution of the observation, i.e., using a smaller beam. Therefore, we further investigated the SCUPOL polarization, which was measured at a similar wavelength as Planck but has a much smaller beam width of $\sim10''$. As seen in Fig.~\ref{fig.bfields}, the magnetic fields ($\sim10''$) inferred from SCUPOL polarization are less regular. The polarization fraction is distributed in the range of 0 - 5\%, approximately, as seen in Figs.~\ref{fig.p_planck} and ~\ref{fig.p_dis_planck}. To compare it with Planck polarization, we further smoothed it to $\sim5'$, and the results can be found in the Appendix.~\ref{app:sm}. Similar to Planck, the SCUPOL measurements reveal a twisted morphology of magnetic fields near IRS 5. If the Planck data was significantly contaminated with noises, the two independent observations are not expected to yield consistent results of magnetic fields in the same area. Therefore, we expect the Planck data to be still reliable in reflecting the magnetic field structure in the L1551 region and thus have kept the Planck data in our analysis.

We further synergized the HAWC+ data obtained most recently with Planck and SCUPOL to investigate the magnetic fields in L1551. Planck and SCUPOL polarization are measured at 850 $\mu$m, which is usually associated with cold, dense gas. HAWC+ provides measurements at a shorter wavelength of 214 $\mu$m, which is linked to gas with a higher temperature \citep{1987stme.book.....H}. As seen in Fig.~\ref{fig.bfields}, the magnetic fields inferred from HAWC+ are still turbulent, but do not appear to be similar to the other two polarization measurements.  While the polarization fraction values at most pixels are no larger than 5\%, the maximum value could be up to more than 20\%, as shown in  Figs.~\ref{fig.p_planck} and \ref{fig.p_dis_planck}. The uncertainties of the HAWC+ polarization are included in the Appendix~\ref{app:pol}. This suggests that the magnetic field morphology may vary in different gas phases (or temperatures) in L1551. While polarization fraction is irrelevant to the following analysis with VGT, it is a significant intrinsic property of polarization measurements and can also shed light on the dust grain alignment within this cloud, which will be discussed in Sec.~\ref{sec:dis}.

\begin{figure}
\label{fig.p_dis_hc+}
    \centering
    \includegraphics[width=1\linewidth]{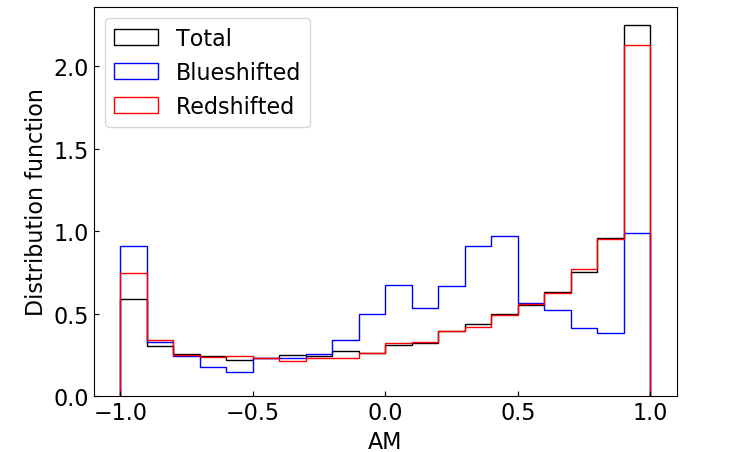}
    \caption{Histogram of AM between Planck and VGT measurements on total $^{12}$CO emission (black line), redshifted emission (red line), and blueshifted emission (blue line).}
\end{figure}

\subsection{Magnetic fields inferred from VGT}
\begin{figure}
\label{fig.s-hc+_dis}
    \centering
    \includegraphics[width=1\linewidth]{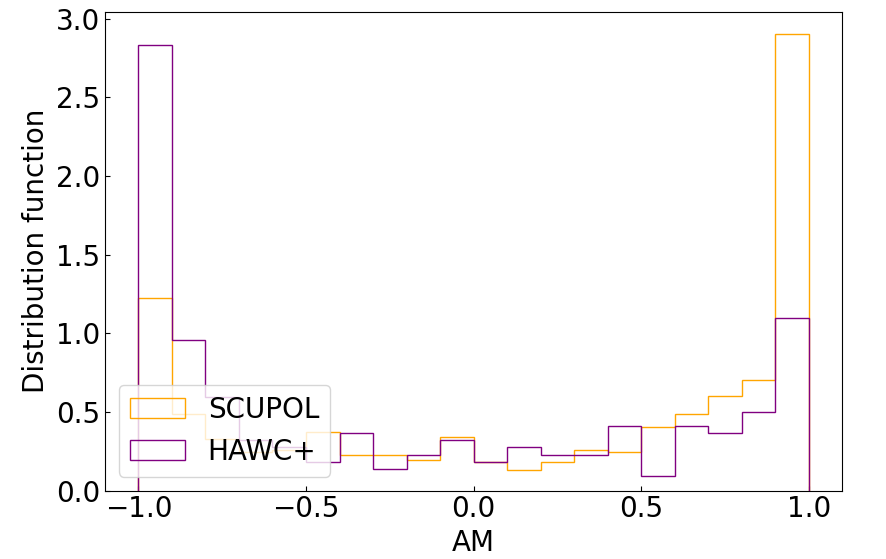}
    \caption{Histogram of AM between VGT measurements on total $^{12}$CO emission and SCUPOL (orange line) and HAWC+ (purple line).}
\end{figure}

Fig.~\ref{fig.p_hc+} shows the magnetic fields inferred from VGT using the $^{12}$CO emission line. The VGT-$^{12}$CO approach uniquely permits the separation of magnetic fields correlated with distinct cloud structures: 
(1) \textit{total}: the full $^{12}$CO emission, integrated from -15 $\rm km$ $\rm s^{-1}$ to 25 $\rm km$ $\rm s^{-1}$; (2) \textit{redshifted}: exclusively for the redshifted component in the gas emission, spanning 7.9 $\rm km$ $\rm s^{-1}$ to 17.4 $\rm km$ $\rm s^{-1}$; (3) \textit{blueshifted}: exclusively for the blueshifted gas, ranging from -3.5 $\rm km$ $\rm s^{-1}$ to 5.5 $\rm km$ $\rm s^{-1}$. The effective resolution of these VGT-$^{12}$CO measurements is 20$\times$20 pixels.

In the fully integrated VGT-$^{12}$CO measurements, the magnetic fields broadly align with the $^{12}$CO intensity configurations in both the northeastern and southwestern parts. Remarkably, the magnetic fields in VGT-$^{12}$CO are not twisted near the IRS 5 protostar. This stands as different from the Planck polarization. As for the red-shifted and blue-shifted ones, the magnetic fields are less regular but still generally follow the gas structures, except for the red-shifted west tail. The magnetic fields are also twisted towards the center around IRS 5 in both red-shifted and blue-shifted cases.

To quantify the agreement between VGT-$^{12}$CO measurements and polarization, we utilize the Alignment Measure (AM; \citealt{2017ApJ...835...41G}), expressed as:

\begin{equation}
\begin{aligned}
{\rm AM} = 2 (\cos^2\theta_{\rm r} - \frac{1}{2})
\end{aligned}
\end{equation}

Here, $\theta_{\rm r} = \vert\phi_{\rm B}-\psi_{\rm B}\vert$. An AM value of 1 implies parallel alignment of $\phi_{\rm B}$ and $\psi_{\rm B}$, while -1 indicates perpendicularity.

Fig.~\ref{fig.I-P_s} shows the spatial distribution of the AM between the fully integrated VGT-$^{12}$CO measurement and the magnetic field orientation inferred from Planck polarization. In the vicinity of the IRS 5 protostar, anti-alignments (AM = -1) are observed. This is also seen in SCUPOL and HAWC+. We attribute this to the strong molecular outflows characteristic of this area. As delineated in Sec.~\ref{sec:method}, velocity gradients stemming from turbulent gas motion are typically perpendicular to magnetic fields and are used in VGT to trace the magnetic fields. This type of gradient is denoted as turbulence-asscociated velocity gradients. However, vigorous outflow dynamics can create pronounced velocity differences, or non-turbulence-associated velocity gradients, which we anticipate to align parallel to the magnetic fields in the presence of substantial outflows.

When extracting gradients from observational data, we inherently deal with the combined velocity field (turbulence plus non-turbulence). Consequently, the VGT-$^{12}$CO — which rotates the velocity gradients by 90 degrees to estimate magnetic field orientations — may result in a perpendicular anti-alignment with polarization when outflows predominate. Moving outward from the center, where outflow influence wanes and turbulence takes precedence, the VGT-$^{12}$CO measurements begin to align with magnetic field orientations inferred from polarization, indicating the relative increase in the significance of turbulent motions.

Furthermore, Fig.~\ref{fig.p_dis_hc+} shows the histogram of the VGT-Planck AM for the red-shifted component, the blue-shifted component and the total gas emission. An intriguing feature is that in both redshifted gas and total emission, the VGT-Planck alignments (AM = 1) are statistically more significant, while in the blue-shifted component, AM concentrates more on $\sim0.5$. We expect this is due to the Planck-inferred magnetic fields do not follow the blueshifted outflows, as seen in Fig.~\ref{fig.bfields}.

\begin{figure*}
\label{fig.s-hc+}
    \centering
    \includegraphics[width=1\linewidth]{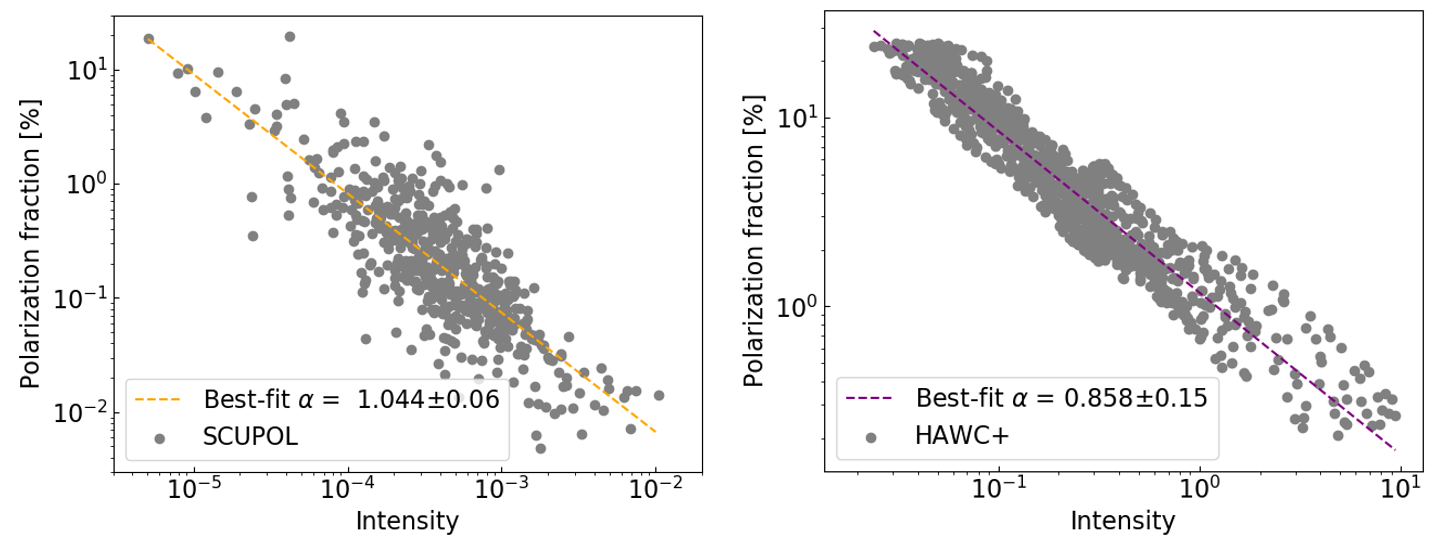}
    \caption{Polarization fraction $p$ as a function of intensity (Stokes $I$) of the SCUPOL (left) and HAWC+ (right) data in the log-log scale. Gray dots represent original data points, and colored dotted lines mark the best-fit trend lines with slopes indicated, which correspond to the $\alpha$ index that reflects the efficiency of dust grain alignment with magnetic fields.}
\end{figure*}

The AM map between the fully integrated VGT-$^{12}$CO measurement and both SCUPOL and HAWC + results has been studied, as seen in Fig.~\ref{fig.s-hc+_dis}. Since the polarimetry surveys observed the protostar systems in L1551 at close-in scales, the magnetic fields they trace are likely related to the protostellar outflows. However, because of the limited sampling ranges of SCUPOL and HAWC+, their measurements are not compared with the magnetic fields associated with individual outflow components. Similarly to the VGT-Planck AM map, anti-alignments of magnetic field measurements inferred by the VGT and the two polarizations are mostly detected near the IRS 5 protostar, suggesting that strong outflows exist in the area, resulting in a difference between VGT and the polarization measurements. Moreover, the anti-alignment is observed in the three polarization measurements with different beam resolutions of $\sim5', 10'',$ and $18.2'$. Since the trend of negative AM was observed in both comparisons with the low-resolution Planck data and the high-resolution SCUPOL and HAWC+, we infer that this may imply the outflow motions dominate the gas dynamics of the L1551 star-forming region starting from $\sim5'$, corresponding to a physical scale of $\sim0.2$~pc. Nonetheless, other observations are needed for cross-checking to confirm this speculation.

\section{Discussion}
\label{sec:dis}
\subsection{Uncertainty}
\subsubsection{Dust grain alignment in the presence of outflows}
The polarization fraction is usually inversely related to the Stokes $I$ in a power-law fashion: $p \propto I^{-\alpha}$. Index $\alpha$ is usually used to assess the effectiveness of dust grain alignment in star-forming regions. It is believed that a value of $\alpha$ = 0 implies perfect alignment between the grains and the magnetic field, while an index of $\alpha=1$ suggests a total loss of alignment \citep{2008ApJ...674..304W,2014A&A...569L...1A,2015AJ....149...31J}. In Fig.~\ref{fig.s-hc+}, we present the correlation between $p$ and $I$ for SCUPOL and HAWC+ data. We find the index is $\sim1.044\pm0.06$ and $\sim0.858\pm0.15$ for SCUPOL and HAWC+, respectively, indicating a lack of grain alignment in the L1551 region. This, along with the negative polarization fractions in Planck results, induced us to provide an additional analysis, in which we observed similarities between the SCUPOL and Planck measurements in vital statistics. As shown in Appendix~\ref{app:sm}, Planck, and SCUPOL, which are nearly at identical wavelengths ($\sim$ 850 $\mu$m) with different spatial resolutions indicate similar magnetic field orientations. While the magnetic field structure inferred from HAWC+ at 214 $\mu$m shows differences with the former two, all three polarization measurements have negative AM with VGT in the central star-forming region. Should dust grain alignment fail to give reliable indications of magnetic fields, the polarization measurements are not expected to show similar magnetic field orientation as well as AM values. Since three independent polarization measurements give similar results, we expect the uncertainties in polarization to be insignificant and that the comparison with VGT stands valid.

\subsubsection{VGT} 
The magnetic field mapped by VGT is subject to two sources of uncertainty: (1) systematic errors in the observational cube and (2) the uncertainty inherent in the VGT algorithm itself. The latter uncertainty arises from the subblock-averaging approach utilized in the algorithm. Specifically, the VGT fits a Gaussian histogram to the orientation of the gradient within a subregion and outputs the angle corresponding to the peak value of the histogram. The associated uncertainty can be quantified as the error $\sigma_{\psi_{\rm s}}(x,y)$ from the Gaussian fitting algorithm within a 95\% confidence level. The uncertainty maps of the VGT-$^{12}$CO measurement are given in the Appendix~\ref{app:vgt}, indicating a globally low level of error.

\subsection{Velocity gradients in the presence of outflows}
When assessing to what extent VGT-$^{12}$CO aligns with the polarization measurements, a common trend is revealed that in the outskirts of the L1551 region, these measurements tend to align with each other reasonably well$\ $(AM $\to$ 1); yet approaching the center of the star-forming region, especially near the protostellar system IRS 5 which is believed to be the major powering source of outflows in L1551 \citep{1998ApJ...499L..75F,2000PASJ...52...81I,2003ApJ...586L.137R}, these measurements seem to produce mostly perpendicular orientations at the same position (AM $\to$ -1). As detailed in Sec.~\ref{sec:res}, this could be attributed to the presence of strong protostellar outflows in such regions. Energetic outflows ejected from the protostar cores generate an outbound acceleration that induces significant velocity gradients. At positions close to the protostars, such non-turbulence-associated gradients generated by the outflow gas motions are the greatest and thus dominate over the gradients induced by turbulent motions in the gas, causing the VGT-inferred magnetic fields to appear perpendicular to the actual magnetic field orientations, which eventually are reflected by the anti-alignments with polarization measurements. Such dominance of non-turbulence-associated gradients stays profound starting from $\sim$ 0.20~pc. At the outer edge away from the protostellar systems, the influence of outflows decreases and turbulence-induced gradients become dominant again. Thus, a high degree of alignment in the VGT and polarization is observed in the outskirts of L1551. This emphasizes the fact that VGT and polarimetry measurements respond differently to the presence of outflows. While polarization could be altered by rotational disruptions of dust grains by both RATs and mechanical torques (METs; \citealt{2022AJ....164..248H}), velocity gradients are closely associated with the kinematic properties of the cloud and thus are mainly affected by the dynamical process. More efforts should be dedicated to investigating in detail what effects those factors, as well as other vital physical properties such as radiations and shocks, have on polarization and gradients, as that can not only better determine the scenario where each technique becomes the most effective, but also promote our insights on the fundamentals of star formation.

\subsection{Prospects of VGT}
In this work, magnetic fields directly associated with the protostellar outflows in the L1551 star-forming region have been investigated with VGT. This novel technique utilizes velocity information from spectroscopic observations to access magnetic fields, hence is capable of separating magnetic fields associated with red-shifted and blue-shifted gas with specific velocities.

This way of applying VGT was inspired by \cite{2022MNRAS.510.4952L} and \cite{2022ApJ...941...92H}, where the technique was used to decompose the magnetic field associated with different velocities components in the supernova remnant W44 and the Central Molecular Zone, respectively. This capacity gives VGT unprecedented advantages in studying astrophysical systems that have complicated velocity components. For instance, as proposed in \cite{2023MNRAS.519.1068L}, different parts of a galaxy, such as spiral arms, typically rotate with different velocities. With VGT, it is possible to isolate them in a velocity space and investigate their magnetic properties individually.

\section{Conclusions}
\label{sec:con}
Magnetic fields are integral to the process of star formation, though tracing them in regions with outflow feedback presents complexities. In this work, we use polarization observations from the Planck at 849 $\mu$m, the SOFIA/HAWC+ measurement at 214 $\mu$m, and the JCMT/SCUPOL at 850 $\mu$m, as well as VGT with the $^{12}$CO (J = 1-0) emission data, to trace the magnetic fields in the star-forming region L1551. Our major discoveries are:
\begin{enumerate}
    \item Upon targeting the star-forming region L1551, known for its discernible outflows, polarization consistently revealed the magnetic fields to be twisted in the direction of the protostar IRS 5.
    \item By applying VGT to the $^{12}$CO (J = 1-0) emission in specific velocity ranges, we were able to separate magnetic fields that are directly tied to the protostellar outflows
    \item While VGT and polarization showed a general alignment on the outskirts of the L1551 region, a trend of mostly perpendicular relative orientations was observed closer to the center of IRS 5.
    \item The observed perpendicular orientations imply that outflows could be dynamically significant from a scale of around $\sim0.2$~pc. This dynamic significance may be causing a change in the direction of velocity gradients by 90 degrees.
    \item A distinct correlation $p \propto I^{-\alpha}$ was observed between the polarization fraction $p$ and the total intensity $I$. Values for $\alpha$ were found to be $1.044\pm0.06$ for SCUPOL and $0.858\pm0.15$ for HAWC+, suggesting that outflows may play a crucial role in determining the alignment between dust grains and magnetic fields within the L1551 region.
\end{enumerate}

\section*{Acknowledgements}
Y.H. and A.L. acknowledge the support of NASA ATP AAH7546, NSF grants AST 2307840, and ALMA SOSPADA-016. Financial support for this work was provided by NASA through award 09\_0231 issued by the Universities Space Research Association, Inc. (USRA). This work used SDSC Expanse CPU at SDSC through allocations PHY230032, PHY230033, PHY230091, and PHY230105 from the Advanced Cyberinfrastructure Coordination Ecosystem: Services \& Support (ACCESS) program, which is supported by National Science Foundation grants \#2138259, \#2138286, \#2138307, \#2137603, and \#2138296. 

\section*{Data Availability}
The data underlying this article will be shared on reasonable request to the corresponding author.



\bibliographystyle{mnras}
\bibliography{example} 



\appendix
\section{Uncertainty of polarimetry measurements}
\label{app:pol}
\begin{figure}
\label{fig.pe_planck}
    \centering
    \includegraphics[width=1\linewidth]{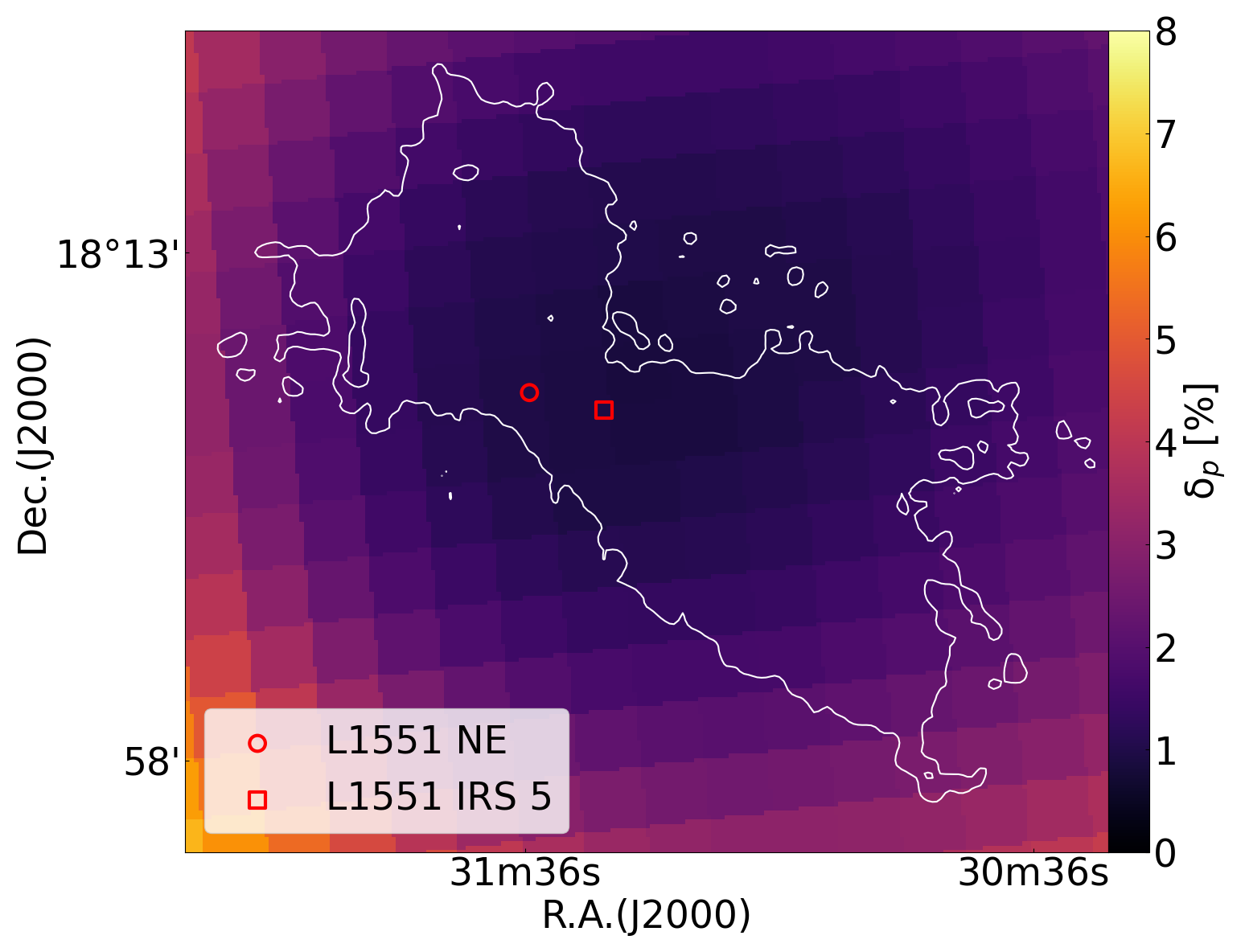}
    \caption{Uncertainty in polarization fraction of the Planck data.}
\end{figure}

\begin{figure}
\label{fig.pe_hc+}
    \centering
    \includegraphics[width=1\linewidth]{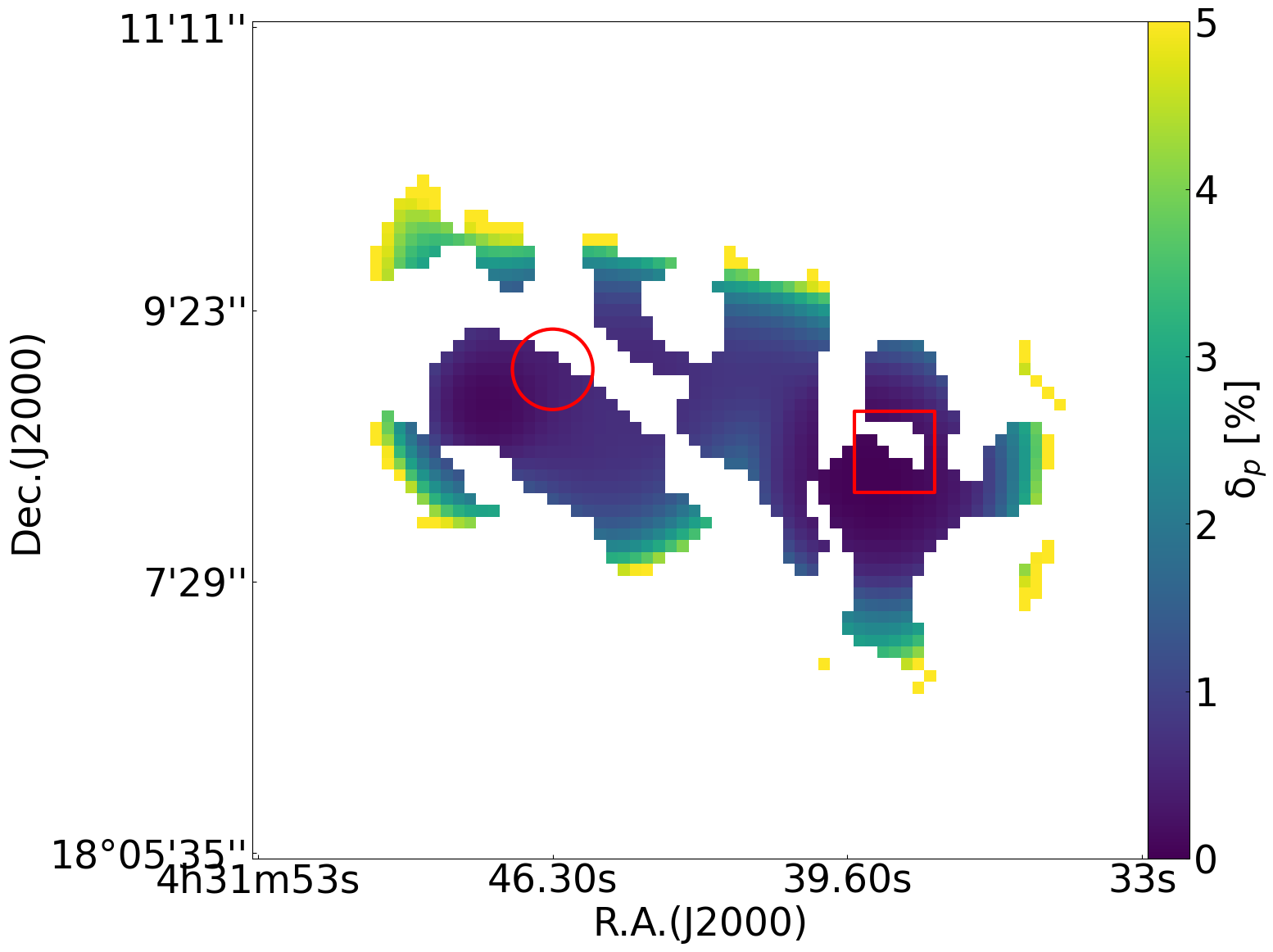}
    \caption{Uncertainty in polarization fraction of the HAWC+ data.}
\end{figure}

\begin{figure}
\label{fig.pa_planck}
    \centering
    \includegraphics[width=1\linewidth]{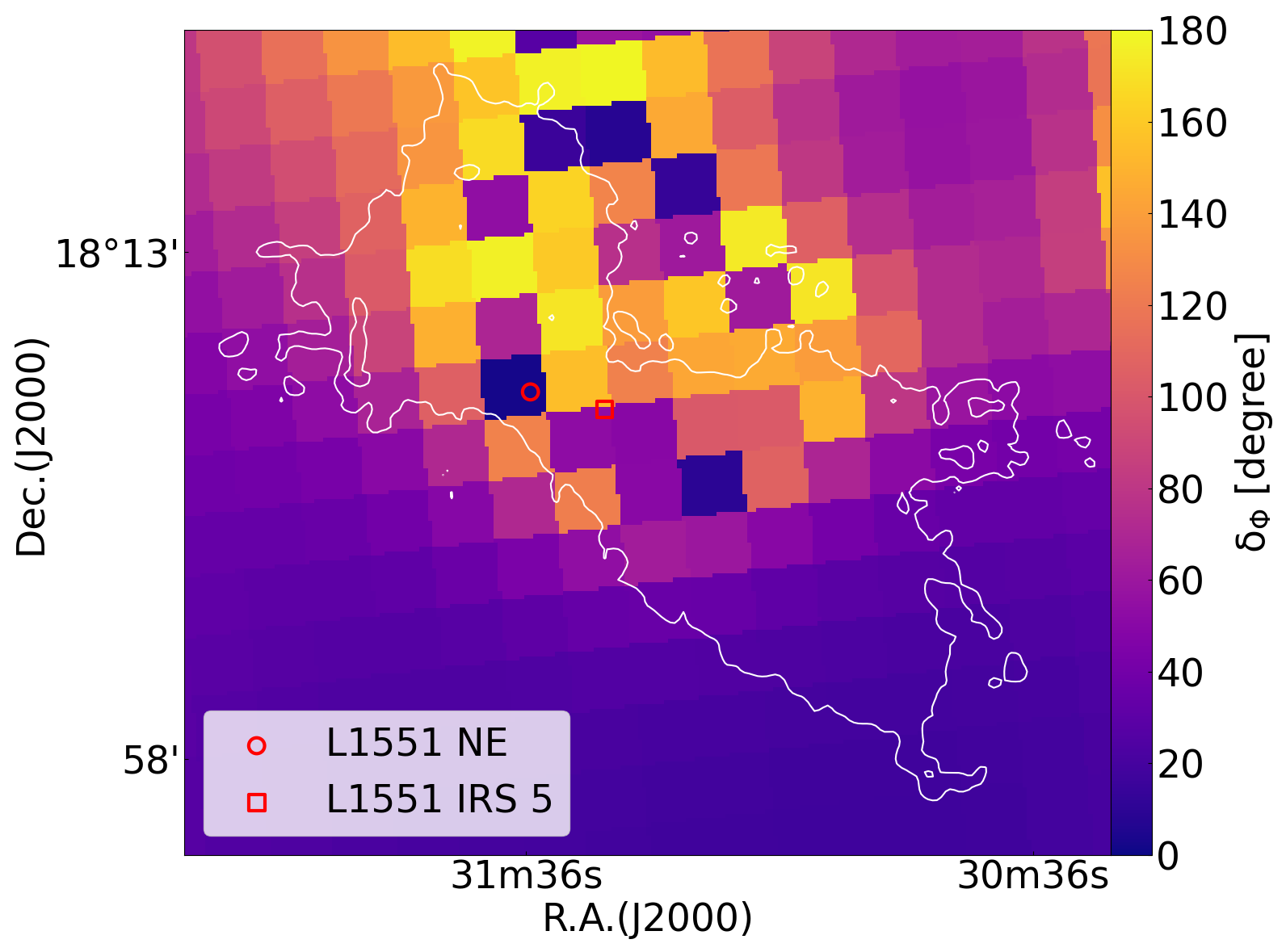}
    \caption{Uncertainty in polarization angle of the Planck data.}
\end{figure}

\begin{figure}
\label{fig.pa_hc+}
    \centering
    \includegraphics[width=1\linewidth]{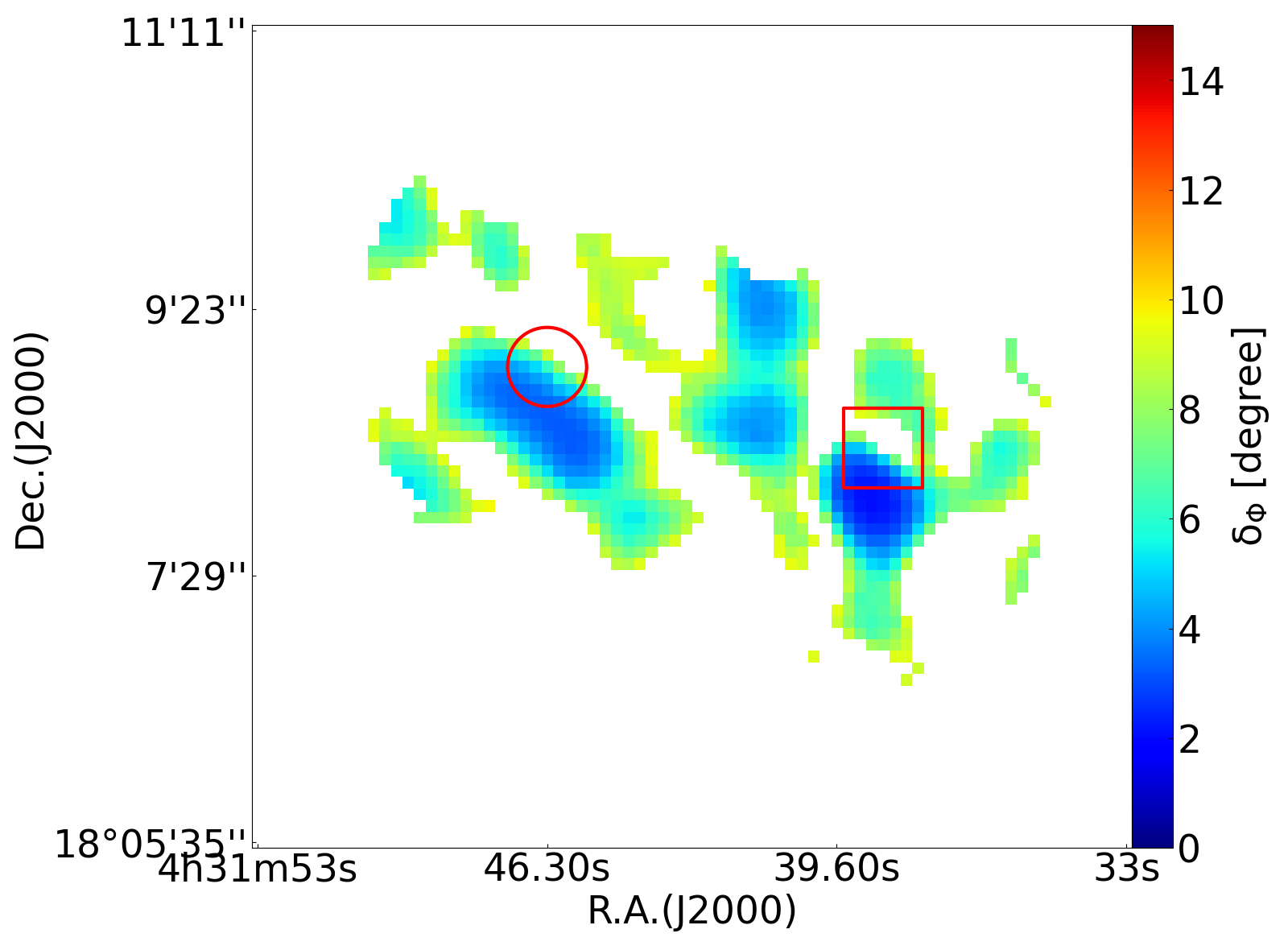}
    \caption{Uncertainty in polarization angle of the HAWC+ data.}
\end{figure}

The uncertainty in the polarization fraction at each pixel of the polarimetry data is calculated as:
\begin{equation}
        \begin{aligned}
           \delta p = \sqrt{\frac{Q^2\delta Q^2+U^2\delta U^2}{I^2(Q^2+U^2)}+\frac{\delta I^2(Q^2 + U^2)}{I^4}},
        \end{aligned}
\end{equation}
$\delta I$, $\delta Q$ and $\delta U$ are the uncertainties in the Stokes parameters. Fig.~\ref{fig.pe_planck}
and Fig.~\ref{fig.pe_hc+} present the uncertainty in polarization fraction of the Planck and HAWC+ data, respectively. Positions of the two protostellar systems, L1551 NE (red circles) and L1551 IRS 5 (red squares), have been highlighted on the graphs. In Planck data, uncertainty remains at low levels ($\leq$ 5\%) in the central area where the main gas intensity structure is located, while higher uncertainties are observed in the outskirt region. A similar trend is also seen in the HAWC+ data, while the uncertainty levels are globally lower than those of Planck.
We also calculated the uncertainty in the polarization angle $\delta \phi$ per pixel as:
\begin{equation}
        \begin{aligned}
           \delta \phi = \frac{1}{2} \sqrt{\frac{Q^2\delta Q^2+U^2\delta U^2}{(Q^2+U^2)^2}},
        \end{aligned}
\end{equation}
Fig.~\ref{fig.pa_planck} and Fig.~\ref{fig.pa_hc+} show uncertainties in polarization angle in the Planck and HAWC+ data. Positions of L1551 NE and L1551 IRS 5 have been highlighted with red circles and red squares, respectively. The uncertainty in Planck becomes significant especially at the northern area, implying strong depolarization and/or poor dust grain alignment efficiency in the region. HAWC+ data suffer less severely from such uncertainties, yet still display an intermediate level of noise (up to $\sim$ 25$^\circ$).

\section{Uncertainty of VGT measurements}
\label{app:vgt}

\begin{figure}
\label{fig.VGTe_t}
    \centering
    \includegraphics[width=1\linewidth]{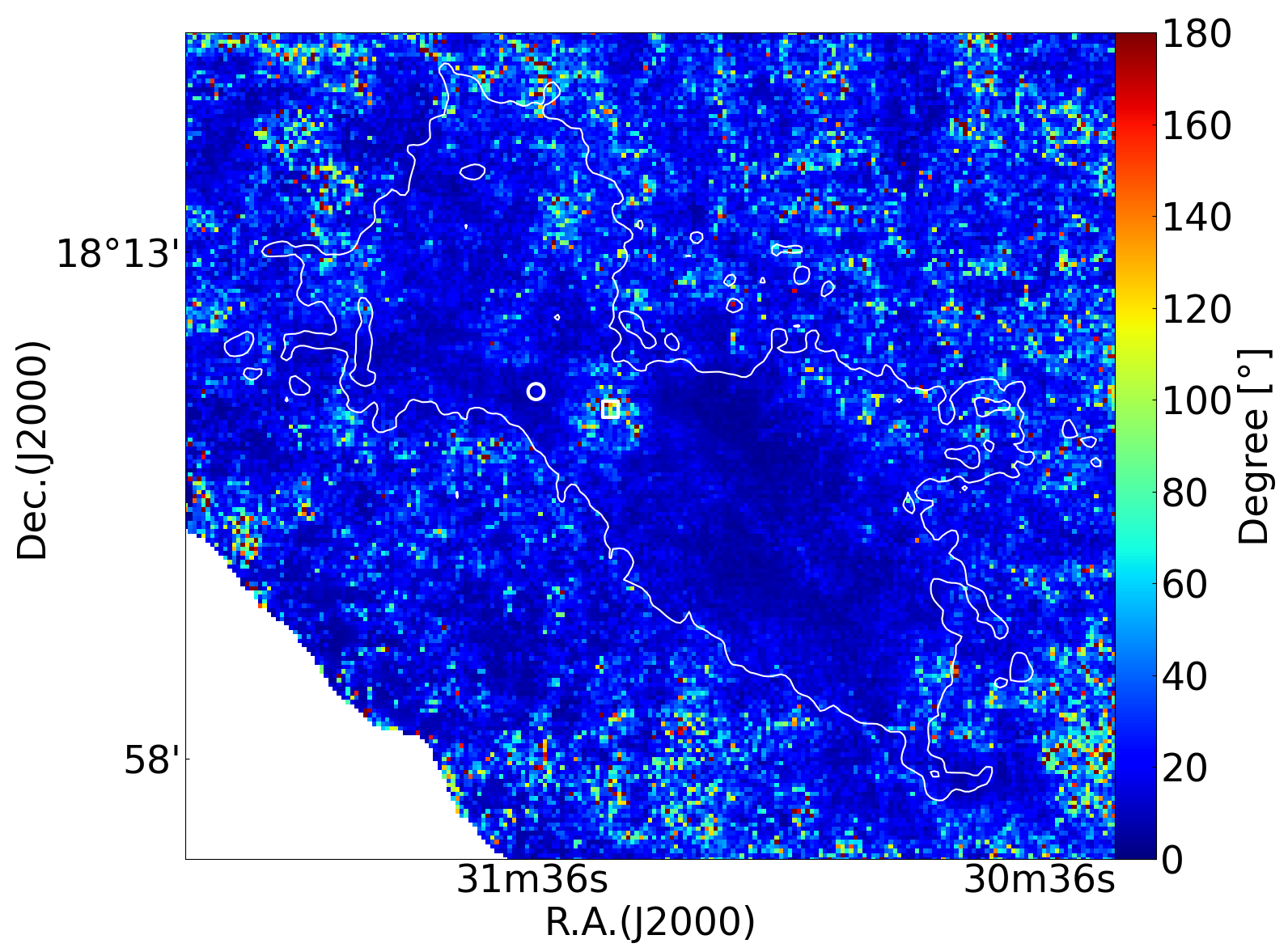}
    \caption{Uncertainty in the VGT-$^{12}$CO measurements.}
\end{figure}

\begin{figure}
\label{fig.VGTe_r}
    \centering
    \includegraphics[width=1\linewidth]{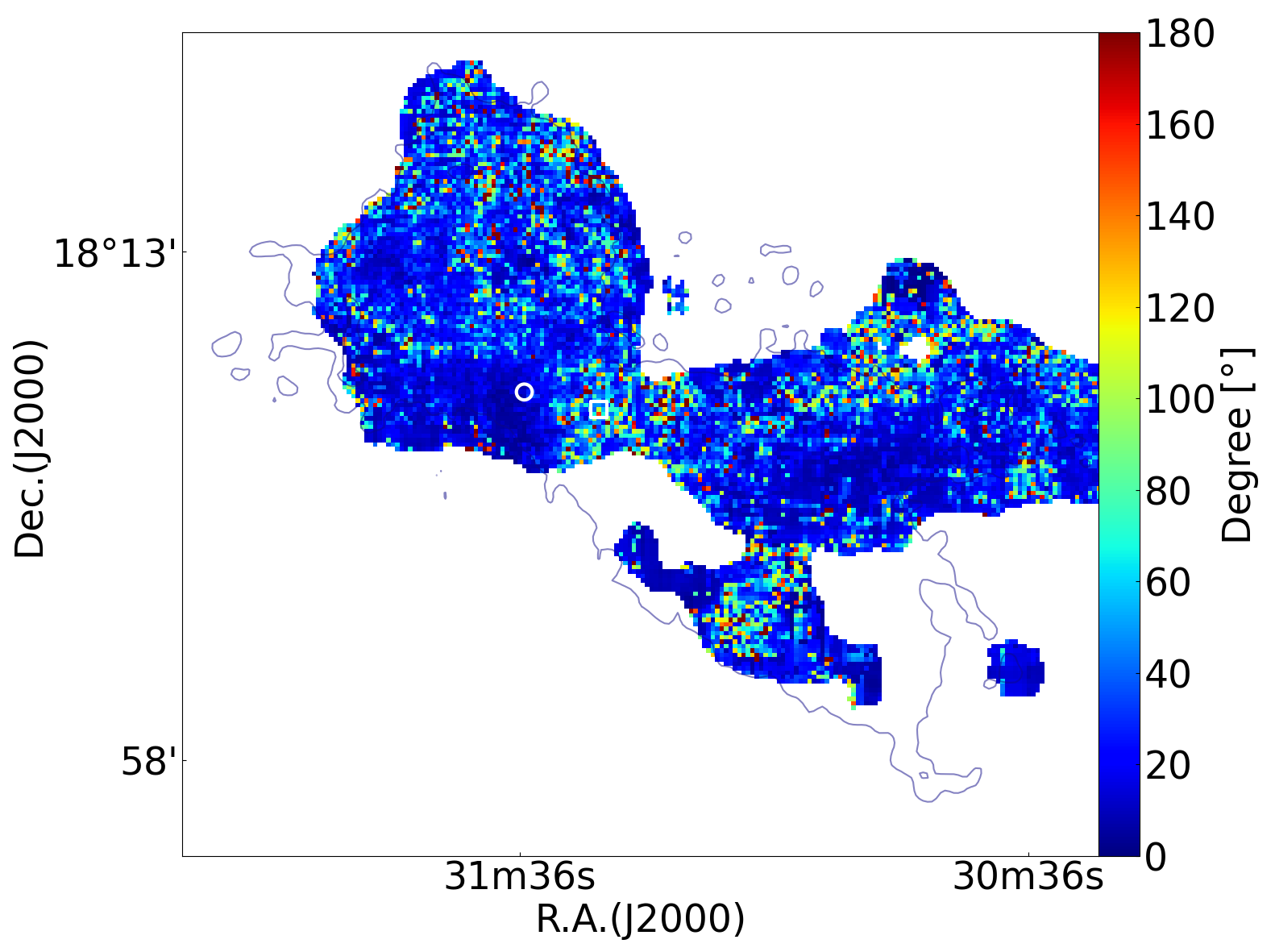}
    \caption{Uncertainty in the VGT measurements on the redshifted $^{12}$CO emission.}
\end{figure}

\begin{figure}
\label{fig.VGTe_b}
    \centering
    \includegraphics[width=1\linewidth]{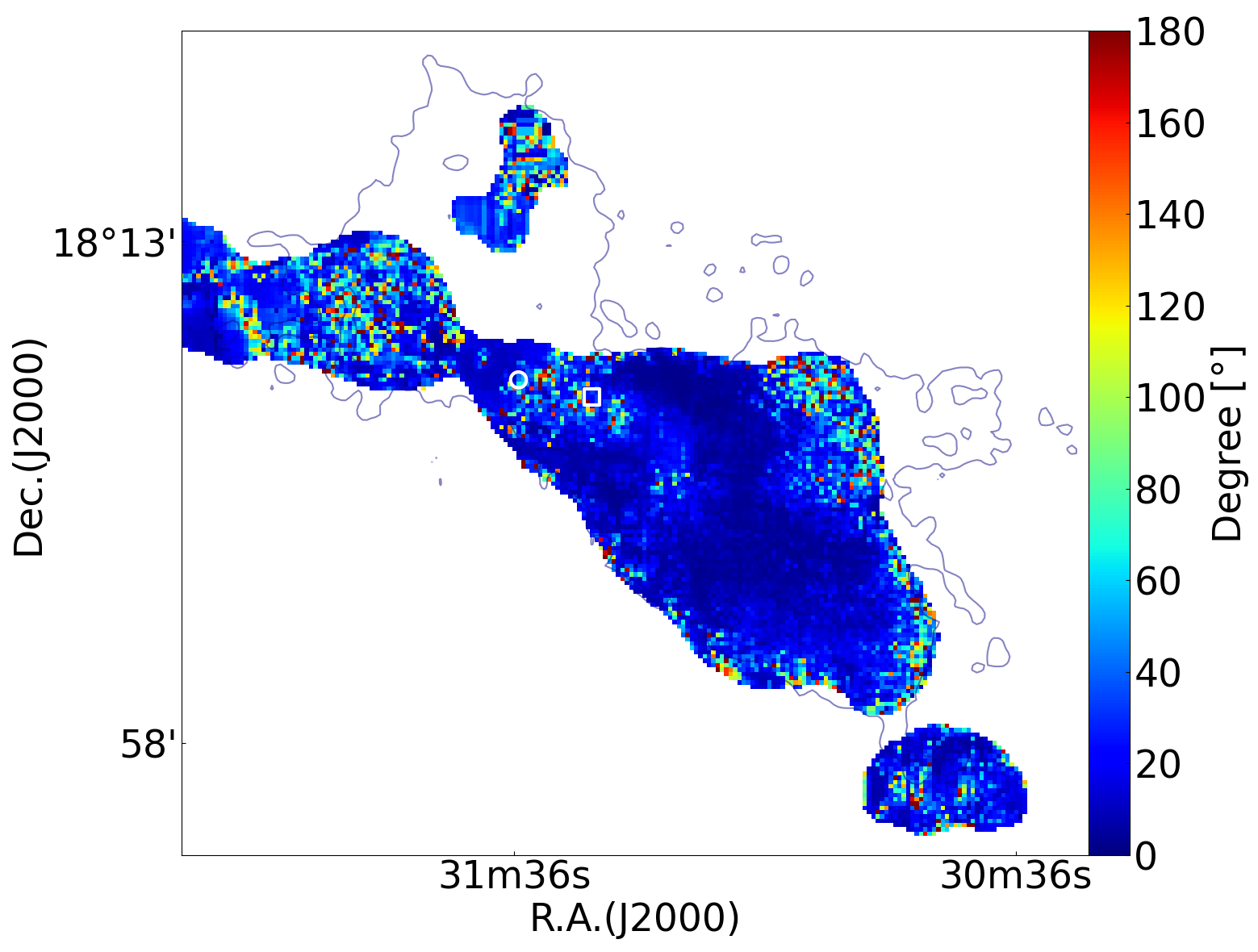}
    \caption{Uncertainty in the VGT measurements on the blueshifted $^{12}$CO emission.}
\end{figure}


The uncertainty in the VGT measurements $\sigma_{\psi_{g}} (x, y, v)$ are obtained along with other systematic errors induced by the VGT algorithm as such:
\begin{equation}
\begin{aligned}
    &\sigma_{{\rm cos}} (x,y,v)  =  |2 {\rm sin}(2 \psi_{\rm gs} (x,y,z)) \sigma_{{\psi_{\rm gs}}} (x,y,v)|\\
    &\sigma_{{\rm sin}} (x,y,v)  =  |2 {\rm cos}(2 \psi_{\rm gs} (x,y,z)) \sigma_{{\psi_{\rm gs}}} (x,y,v)|\\
    &\sigma_{q} (x,y,v)  =  |{\rm Ch} \cdot {\rm cos(2 \psi_{\rm gs})}|\sqrt{(\sigma_{n}/{\rm Ch})^2+(\sigma_{{\rm cos}}/{\rm cos(2 \psi_{\rm g})})^2}\\
    &\sigma_{u} (x,y,v)  =  |{\rm Ch} \cdot {\rm sin(2 \psi_{\rm gs})}|\sqrt{(\sigma_{n}/{\rm Ch})^2+(\sigma_{{\rm sin}}/{\rm sin(2 \psi_{\rm g})})^2}\\
    &\sigma_{Q} (x,y)  =  \sqrt{\sum_{v} {\sigma_{q}(x,y,v)^2}}\\
    &\sigma_{U} (x,y)  =  \sqrt{\sum_{v} {\sigma_{u}(x,y,v)^2}}\\
    &\sigma_{\psi_{g}} (x,y) = \frac{|U_{\rm g}/Q_{\rm g}|\sqrt{(\sigma_{Q}/Q_g)^2+(\sigma_{U}/U_g)^2}}{2[1+(U_{\rm g}/Q_{\rm g})^2]}
\end{aligned}
\end{equation}
here, $\sigma_{\psi_g}$ denotes the angular uncertainty of the magnetic field measurements, $\sigma_n (x, y, v)$ represents the noise and propagated error in thin velocity channel Ch$(x, y, v)$, and $\sigma_Q (x, y)$, $\sigma_U(x, y)$ account for the respective uncertainty of the pseudo-Stokes parameters $Q_g (x, y)$, $U_g (x, y)$.

Fig.~\ref{fig.VGTe_t}, Fig.~\ref{fig.VGTe_r} and Fig.~\ref{fig.VGTe_b} present uncertainties in the VGT measurements on the total $^{12}$CO emission, redshifted emission, and blueshifted emission, respectively. Positions of L1551 NE and L1551 IRS 5 have been highlighted with white circles and white squares, respectively. All these measurements indicate reasonably a low level of error, despite of some random fluctuations.

\section{Supplemental details}
\label{app:sm}

\begin{figure}
\label{fig.co_spectrum}
    \centering
    \includegraphics[width=1\linewidth]{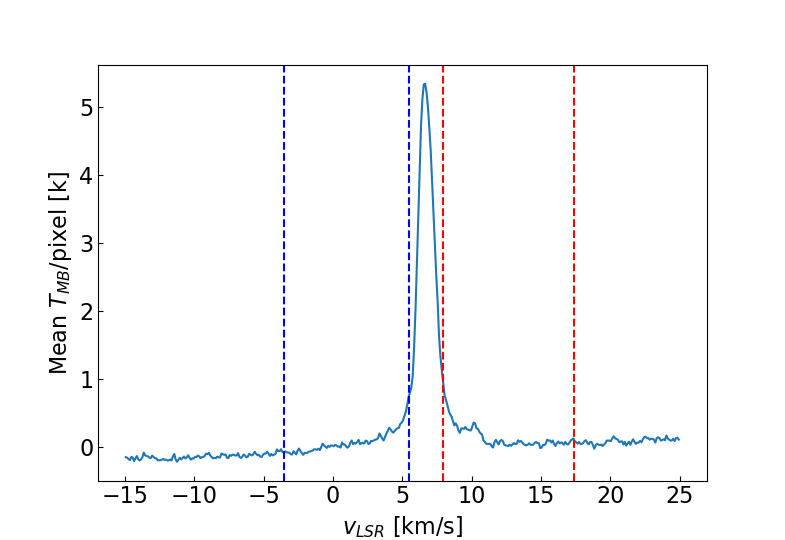}
    \caption{Spectral line of $^{12}$CO emission. Red and blue dotted lines mark the redshifted and blueshifted regions, respectively.}
\end{figure}

\begin{figure}
\label{fig.s-unsmooth}
    \centering
    \includegraphics[width=1\linewidth]{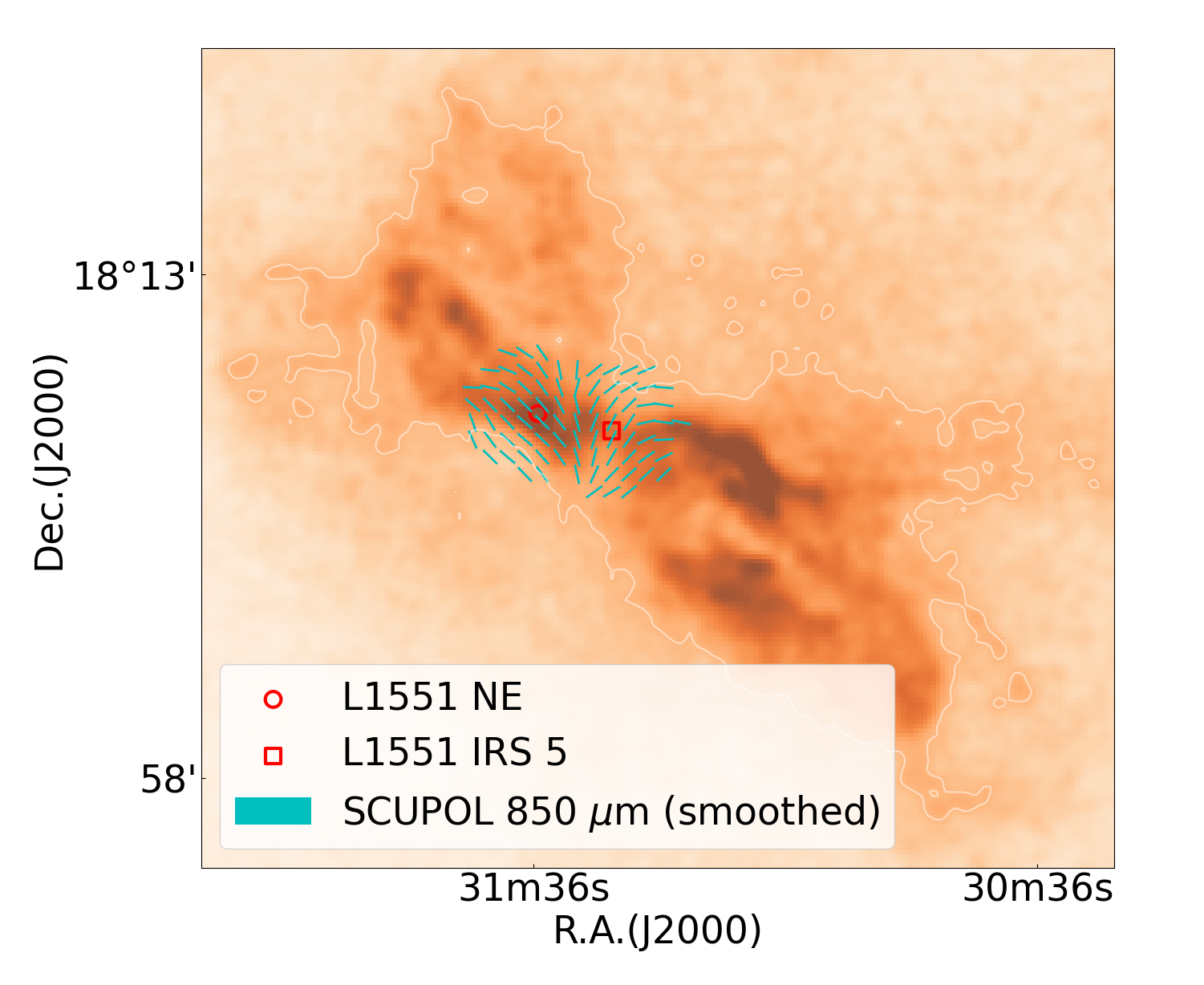}
    \caption{Smoothed magnetic field orientation map inferred from the SCUPOL data.}
\end{figure}

\begin{figure}
\label{fig.hc+-unsmooth}
    \centering
    \includegraphics[width=1\linewidth]{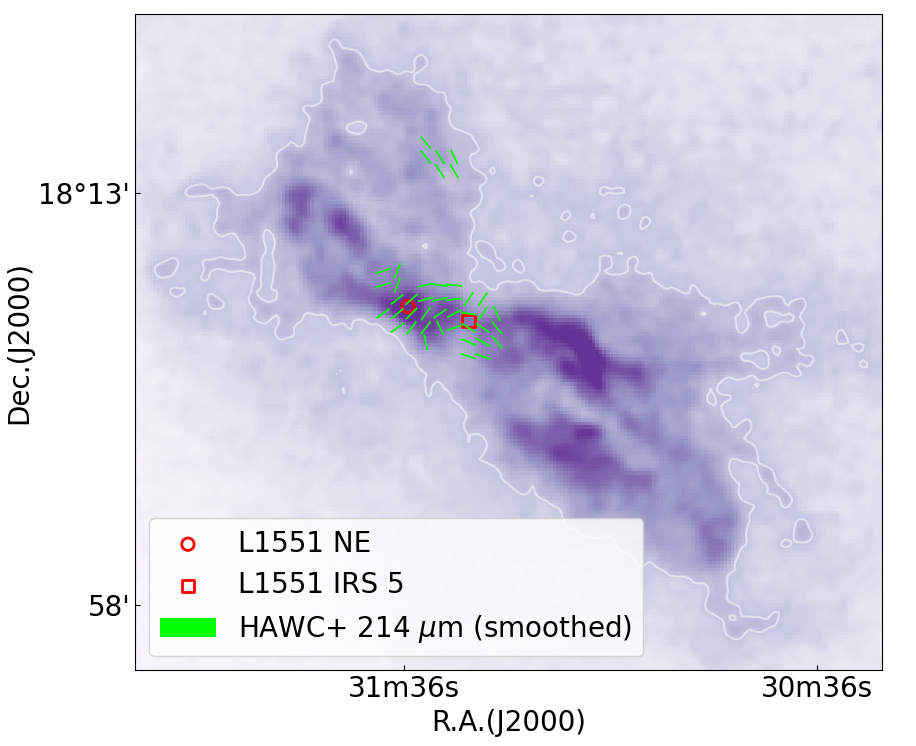}
    \caption{Smoothed magnetic field orientation map inferred from the HAWC+ data.}
\end{figure}

\begin{figure}
\label{fig.pa_hist}
    \centering
    \includegraphics[width=1\linewidth]{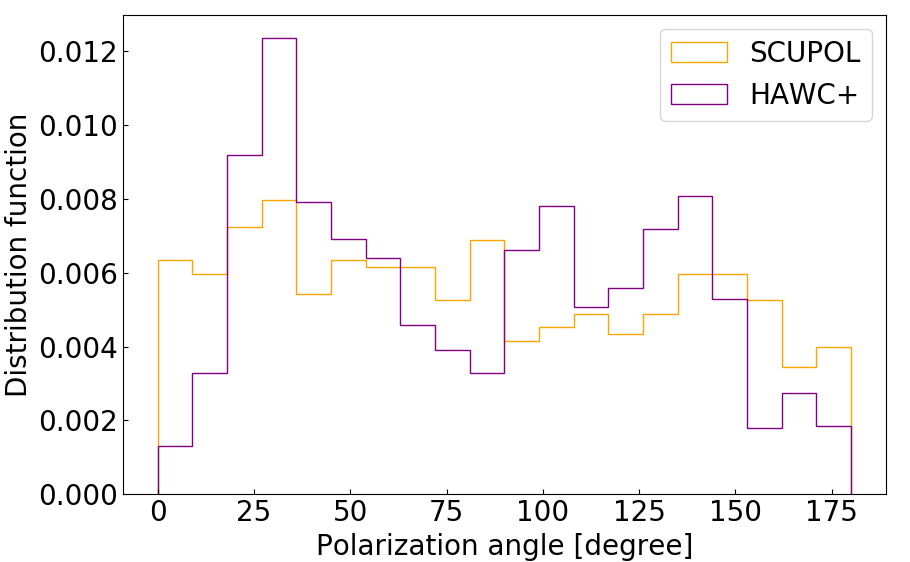}
    \caption{Histogram of polarization angle in the SCUPOL (orange line) and HAWC+ (purple line) data. Note that the angle is defined in the IAU convention.}
\end{figure}

\begin{figure}
\label{fig.i-p_pl}
    \centering
    \includegraphics[width=1\linewidth]{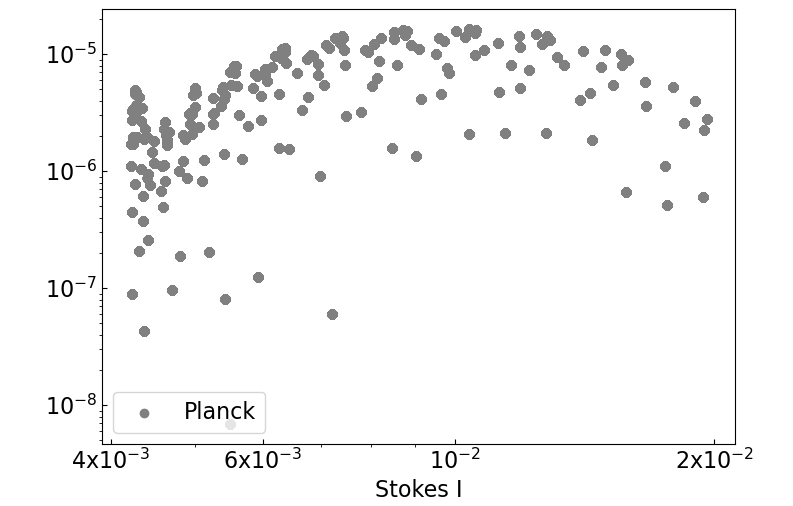}
    \caption{Polarization fraction $p$ as a function of intensity (Stokes $I$) of the Planck data in the log-log scale. }
\end{figure}
Fig.~\ref{fig.co_spectrum} presents spectrum of the $^{12}$CO emission line. The redshifted and blueshifted components at $v_{LSR} \in$ [7.9, 17.4] $\rm km$ $\rm s^{-1}$ and [-3.5, 5.5] $\rm km$ $\rm s^{-1}$ respectively have been indicated. Fig.~\ref{fig.s-unsmooth} and Fig.~\ref{fig.hc+-unsmooth} are the magnetic field orientation maps inferred from the SCUPOL and HAWC+ data which have been applied a Gaussian filter to smooth the images to 5$'$ so that they are comparable with the Planck data. The smoothed SCUPOL map noticeably differs from the original figure, revealing that in a larger scale, the magnetic fields present a similar trend as in the Planck map that
the orientations bend dramatically near IRS 5. Fig.~\ref{fig.pa_hist} is the histogram of polarization angle (east from the north) in the SCUPOL and HAWC+ data. Fig.~\ref{fig.i-p_pl} shows the I-p correlation of the debiased Planck data of the L1551 region. Its low polarization fraction along with the $\alpha$ index close to 0 imply that it has a questionable accuracy for tracing the dense part of the cloud.

\bsp	
\label{lastpage}
\end{document}